\documentclass[aps,pre,superscriptaddress,groupedaddress]{revtex4-2}  
\usepackage{graphicx}  
\usepackage{dcolumn}   
\usepackage{bm}        
\usepackage{amssymb}   
\usepackage{amsmath}   
\usepackage{mathtools} 
\usepackage{color}
\usepackage{ulem}

\usepackage[colorlinks,linkcolor=blue,urlcolor=blue,citecolor=blue]{hyperref}

\usepackage{subfig}

\begin{document}

\def\be{\begin{equation}}
\def\ee{\end{equation}}
\def\bea{\begin{eqnarray}}
\def\eea{\end{eqnarray}}
\def\l{\label}

\newcommand{\eref}[1]{Eq.~(\ref{#1})}%
\newcommand{\Eref}[1]{Equation~(\ref{#1})}%
\newcommand{\fref}[1]{Fig.~\ref{#1}} %
\newcommand{\Fref}[1]{Figure~\ref{#1}}%
\newcommand{\sref}[1]{Sec.~\ref{#1}}%
\newcommand{\Sref}[1]{Section~\ref{#1}}%
\newcommand{\aref}[1]{Appendix~\ref{#1}}%
\newcommand{\sgn}[1]{\mathrm{sgn}({#1})}%
\newcommand{\erfc}{\mathrm{erfc}}%
\newcommand{\erf}{\mathrm{erf}}%

\title{Singular optimal driving cycles of stochastic pumps}

\author{Ilana Bogod}
\affiliation{Schulich Faculty of Chemistry, Technion - Israel Institute of Technology, Haifa 3200008, Israel}

\author{Saar Rahav}
\email{rahavs@technion.ac.il}
\affiliation{Schulich Faculty of Chemistry, Technion - Israel Institute of Technology, Haifa 3200008, Israel}

\date{\today}

\begin{abstract}
The investigation of optimal processes has a long history in the field of thermodynamics. It is well known that finite-time processes that minimize dissipation
often exhibit discontinuities. We use a combination of numerical and analytical approaches to study 
the  driving cycle that maximizes the output in a simple model of a stochastic pump: a system driven out of equilibrium by a cyclic
variation of external parameters. We find that this optimal solution is singular, with an infinite rate of switching between
sets of parameters. The appearance of such singular solutions in thermodynamic processes is surprising, but we argue that
such solutions are expected to be quite common in models whose dynamics exhibit exponential relaxation, as long as the driving period is not externally fixed, and is allowed to be arbitrarily short.
Our results have implications to artificial molecular motors that are driven in a cyclic manner.
\end{abstract}

\maketitle

\section{Introduction}

Questions regarding the optimal way of performing a task, or operating a machine, fit naturally into the scope of the theory of thermodynamics \cite{Callen}.
Examples range from Carnot's celebrated work on heat engines \cite{Carnot} to more modern geometrical approaches  \cite{Band1982,Sivak2012,Zulkowski2015}. Technological advances allow the manipulation and observation of molecular motors, nanoscale machines, and microscopic information or heat engines. The theory of stochastic thermodynamics offers a consistent and illuminating thermodynamic description
of such small out-of-equilibrium systems \cite{ST-1,ST-2,ST-3,ST-4}. In particular, the theory accounts for the unavoidable fluctuations due to the random
interactions of a small system with its environment. This research effort has led to several fundamental results that enhance our understanding of out-of-equilibrium 
systems and processes, such as fluctuation relations \cite{Evans93,Evans94,Gallavotti95,Seifert05,Jarzynski97,Crooks99},  thermodynamic uncertainty relations \cite{TUR-1,TUR-2,TUR-review,Seifert2019}, and a renewed interest in the thermodynamics of information \cite{Parrondo2015}. The predictions of the theory were tested experimentally using colloidal particles \cite{Carberry2007,Toyabe2010,Blickle2012,Mehl2012,Admon2018}, RNA hairpins \cite{Colin2005,Alemany2012}, single electron boxes \cite{Saira2012,Koski2014}, molecular motors \cite{Hayashi2010,Hayashi2018}, and more \cite{ST-5}.

A substantial body of research was devoted to the investigation of the role  of optimal processes in stochastic thermodynamics. Schmiedl and Seifert  have pointed out that finite time processes which minimize the 
work often exhibit discontinuities at initial and final times \cite{Schmiedl2007}. The appearance of such discontinuities is known in control theory \cite{LiberzonBook}, but
is somewhat surprising in the context of thermodynamic processes. This result therefore generated considerable interest, and similar discontinuities were found in several other out-of-equilibrium systems \cite{Then2008,Esposito2010,Aurell2011}. Several works showed how to apply techniques from control theory to stochastic thermodynamics problems, see e.g. \cite{Aurell2011,Ginanneschi2013,Large2019,Prados2021}. While much of the work on the subject focused on finite time transitions between two states, several papers studied the optimal operation of periodically driven systems \cite{Bauer2016,Takahashi2020,Funo2020,Remlein2021}.

In this work we investigate the optimal driving cycles of stochastic pumps. The term stochastic pump is used to describe models with discrete states which are driven away from 
equilibrium due to cyclic variation of system parameters. Several experimental realizations of artificial molecular machines can be modeled as stochastic pumps \cite{Leigh2003,Browne2006,Erbas2015,Pezzato2017}. Various aspects
of stochastic pumps were studied theoretically in the last two decades. These include: i) the underlying similarities and differences between steady-states and periodically driven systems \cite{Raz2016,Knoch2019}; ii)  adiabatic and geometrical pumping \cite{Astumian2003, Sinitsyn2007,Ohkubo2008,Raz2016b}; the no-pumping theorem \cite{Rahav2008,Chernyak2008,Horowitz2009,Maes2010,Mandal2011,Ren2011,Asban2014,Asban2015}; as well as other aspects of such pumps \cite{Parrondo1998,Sokolov1999,Rahav2011,Chernyak2012,Chatterjee2014,Brandner2015,Proesmans2016,Koyuk2018,Barato2018}. Additional results are summarized in several review papers devoted to this topic \cite{Sinitsyn2009,Astumian2011,Astumian2018,Wang2022}.

The optimization problem studied here exhibits two aspects that should be highlighted. One is the presence of a resisting external force. The pump performs work against this external force. 
Much of the existing literature on optimal stochastic thermodynamic processes was centered on minimizing dissipation when transitioning between two states, or maximizing a pump current. By focusing on the
pump's power output, we set up a  different optimization problem.
The second aspect is the full freedom of choosing the time dependence of parameters (within a predefined range). Specifically, there are no additional restrictions on the shape of the driving cycle or its period.
This in turn means that there are no evident approximations that can be used to reduce the dimension of the optimization problem. Nevertheless, progress can be made by considering a simple two-site model of a stochastic pump. We use a combination of analytical and numerical techniques
to identify the driving cycles that maximize the power output. Interestingly, these optimal cycles  turn out to be singular. Specifically, the external parameters alternate between two sets of values. Surprisingly, maximum output is found when the rate of switching between the sets approaches infinity. While such solutions are known in the literature on control theory \cite{ChatteringC,SlidingC}, they are uncommon in the context of thermodynamic processes. Our results therefore add a new facet to the study of the stochastic thermodynamics of optimal processes.

Our paper is structured as follows. In Sec. \ref{sec:molmach} we discuss some qualitative aspects of artificial molecular machines that serve as motivation the problem investigated in what follows.
We present the model and define the relevant thermodynamic observables in Sec. \ref{sec:model} . Sec. \ref{sec:limitations} describes 
some physically motivated restrictions that we place on the driving cycles. Such restrictions are needed to make the optimization problem well defined. In Sec. \ref{sec:cycle} we identify and study a family of driving
cycles which are a good candidates for optimal solutions. We then demonstrate in Sec. \ref{sec:stab} that these cycles locally maximize the output of the pump. A numerical approach based
on genetic algorithms is presented and applied in Sec. \ref{sec:num}. It suggests that the optimal solution is attained when the period of the cycles from Sec. \ref{sec:cycle}
approaches 0. In Sec. \ref{sec:eps} we study a regularized problem in which a small cost is associated to changes in site energies. A good agreement between the analytical and numerical approaches is found for this regularized problem. 
Furthermore, the regularized solutions approach the their singular counterparts from Sec. \ref{sec:num} when the additional cost is decreased.
Finally, we discuss the implications and generality of our results in Sec. \ref{sec:disc}.

\section{Motivation: Periodically driven molecular machines}

\label{sec:molmach}

Our bodies are teeming with proteins that act as machines. Life would not have been possible without molecules that transport cargo, copy or translate nucleic acids, assist cell division, and more \cite{Howardbook}. The fascination with molecules that can act as machines  inspired chemists to design and synthesize artificial molecular machines.
 The success of this fascinating research effort has resulted in the awarding of the 2016 Nobel prize in Chemistry, which was awarded to Sauvage, Stoddart, and Feringa.

The work presented below is motivated by a certain type of artificially made molecular machines, built from  complexes of mechanically interlocked molecules~\cite{Pezzato2017}.  As pioneered by the groups of Stoddart and Leigh, such mechanically interlocked molecules can be operated as machines by making a closed cycle of changes in the
molecules or in their environment. The motion of molecules in solution is highly overdamped and exhibits relatively large fluctuations. The dynamics is often described in terms of transitions between a few metastable conformations.  Since the changes made during the cycle  are often local, it is natural to model them theoretically as externally controlled changes in the free energy of the metastable states and barriers between them. When these are varied cyclically, the system is operated as a stochastic pump. 
 Artificial molecular machines of this kind include switches, shuttles, and rotary motors. Specific examples and details regarding the chemical structure and operation can be found in several review papers \cite{Leigh2003,Browne2006,Erbas2015,Pezzato2017,Astumian2018}.

\begin{figure}[htb]
\centering
\subfloat[][]{\includegraphics[width=0.4\textwidth,height=5.1cm]{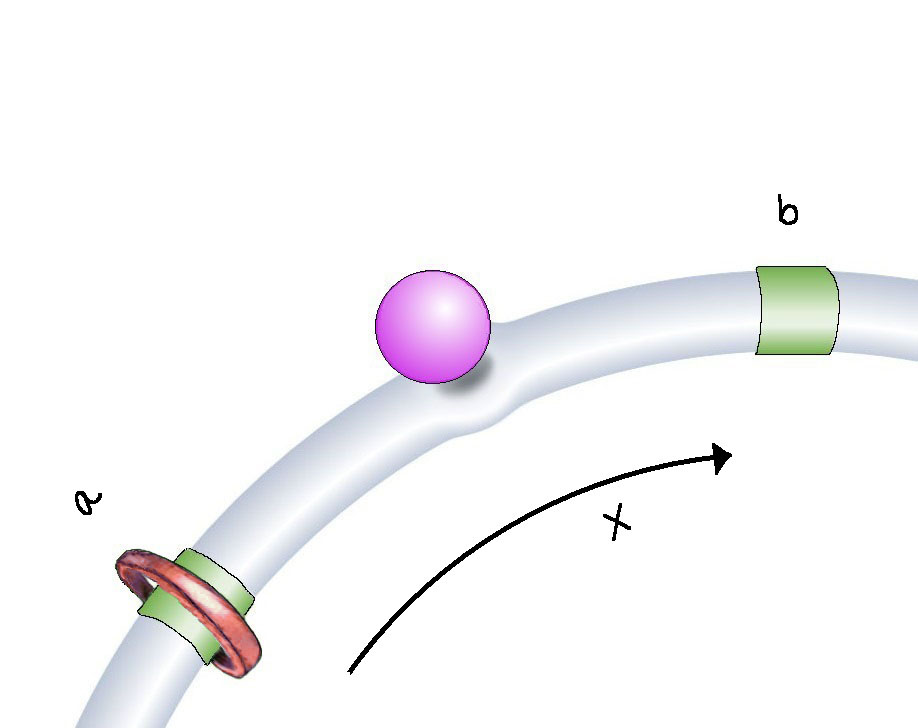}
\label{fig:1a}}
\hspace{8pt}
\subfloat[][]{\includegraphics[width=0.4\textwidth,height=5.1cm]{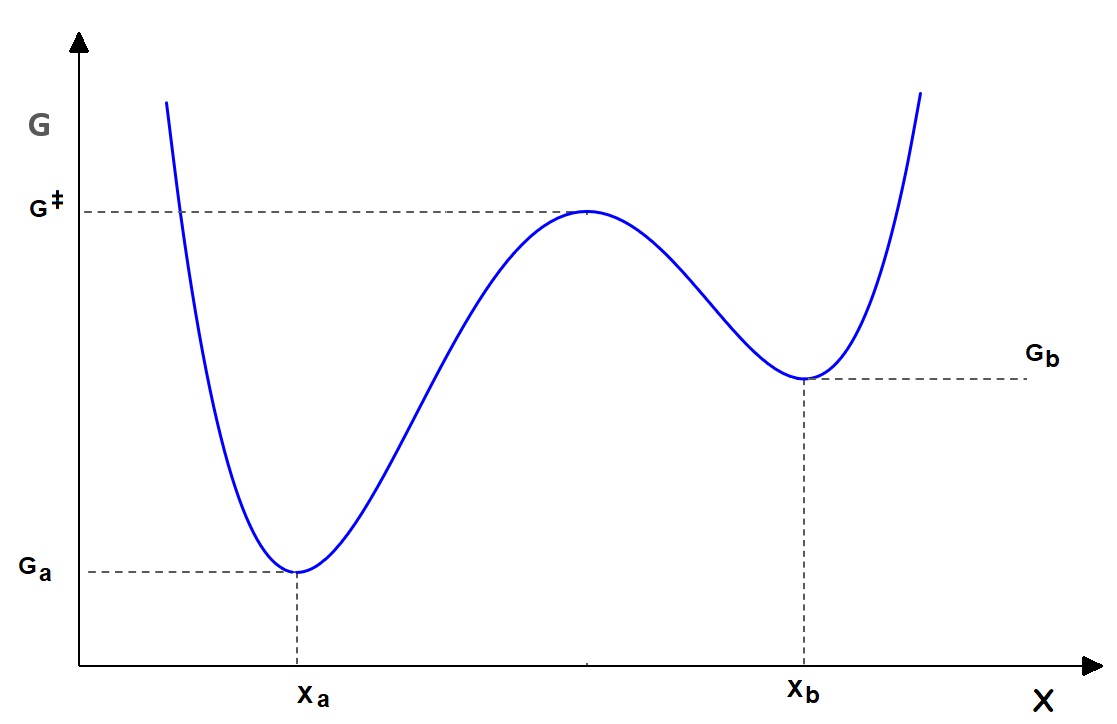}
\label{fig:1b}} \\
\subfloat[][]{\includegraphics[width=0.4\textwidth]{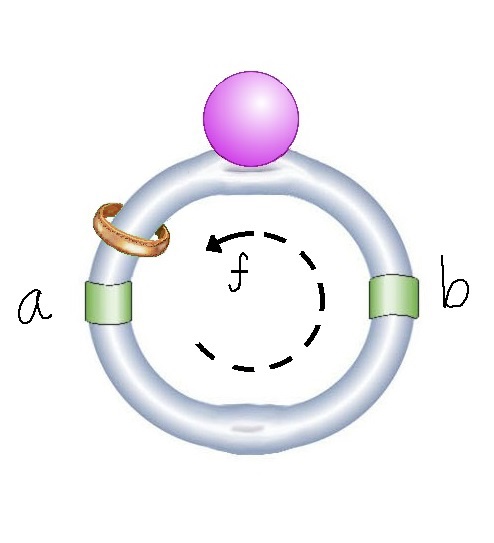}
\label{fig:1c}}
\caption[]{Heuristic depiction of artificial molecular machines that are modeled as stochastic pumps.
\subref{fig:1a} Part of a mechanically interlocked system in which a small ringlike molecule can move along a larger backbone molecule. The two green regions depict areas capable of forming favorable non-covalent interactions between the ring and backbone. The dent and purple ball represent a region with steric or repulsive interactions that acts as a local barrier. (The barrier is lower when the ball is absent.)
\subref{fig:1b} A one-dimensional  free energy landscape can be used to describe the system depicted in Fig. \ref{fig:motivation}(a). The minima represent the landing sites, and the maximum represents the barrier. The free energies of the wells and the barrier can be modified by replacing the side groups attached to the backbone. The system is typically in a solvent at room temperature. As a result transitions between the wells are thermally activated, with rates given by Eq. (\ref{eq:formalrate}). 
\subref{fig:1c} A depiction of the model studied in this work. A small ringlike molecule can move between two stable landing sites on a larger ring. Each of the transition paths has its own barrier.}
\label{fig:motivation}%
\end{figure}

Fig. \ref{fig:motivation}(a)  heuristically depicts a part of a molecular machine of this kind. A small ringlike molecule can move along the backbone of a larger molecule.  $a$ and $b$ are energetically favorable binding sites for the ring, due to hydrogen bonds or van der Waals interactions. The purple ball represents part of the backbone that acts as a barrier.

The free energy landscape of such a system is the multi-dimensional generalization of the simple free energy landscape shown in Fig. \ref{fig:motivation}(b). The motion in such a landscape consists  of relatively rapid local equilibration in the wells, with occasional transitions between the wells. When $G^\ddagger -G_a \gg k_B T$ the transition between the metastable configuration are thermally activated, with rates
\begin{equation}
\label{eq:formalrate}
R_{ba} \simeq \nu \exp \left[-\frac{G^\ddagger -G_a}{k_B T} \right].
\end{equation}
Here $G_a$ and $G^\ddagger$ are the free energies of the configuration $a$ and the ring at the top of the barrier, and $\nu$ is an attempt frequency.
For overdamped dynamics the expression (\ref{eq:formalrate}) can be justified using Kramers rate theory \cite{Hanggi1990}. Such rates are commonly used to model the dynamics of artificial molecular machines.

The existence of this class of periodically driven molecular machines raises various fundamental questions regarding their thermodynamics. What is the optimal way of driving such systems? Should externally controlled parameters be varied gradually, or in an abrupt  manner? Here we study how the output of such systems should be maximized. The number of time-dependent control parameters makes this a nontrivial problem. We therefore focus on a simple example. Consider the system depicted in Fig. \ref{fig:motivation}(c). It has two metastable configurations (which we call sites) and two possible transition between them. We imagine that the free energies of the two sites and the barriers can be manipulated externally. Our model also includes a resisting force $f$ that makes counter-clockwise transitions more likely. (Strictly speaking, $f$ is a shorthand for the force times the distance between sites, and thus has units of energy.) The system is operated as a pump when the ring is driven against this resisting force.

Interestingly, our model is quite similar to a molecular machine realized by the Leigh group \cite{Wilson2016}. Their artificial molecular motor has the structure depicted in  Fig. \ref{fig:motivation}(c), but its kinetics differs from the one we explore in several important aspects. The system realized in \cite{Wilson2016} was autonomous, and furthermore exhibited non-trivial coupling between the position of the smaller ringlike molecule and the kinetics of the barriers. In contrast, we assume that the barriers are externally controlled in  a way that does not depend on the sites. In addition the machine of \cite{Wilson2016} had no resisting force. Nevertheless, this beautiful experiment demonstrates the feasibility of molecular machines of this kind, and motivates the theoretical investigation of their dynamics and thermodynamics.

\section{The two-site model}
\label{sec:model}

The system we examine is a two-site stochastic pump, as depicted in Fig. \ref{fig:1c}.
Each site has an energy that can be manipulated externally. (To simplify the terminology we will refer to the site and barrier free energies as energies in the following.  The two distinct bidirectional transitions have local energy barriers $B_{1,2}$
that must be overcome for a transition to occur. The  resisting  force $f$ makes counter-clockwise transitions more likely by biasing the transitions.
These rates are taken to be Markovian and thermally activated, meaning that the stochastic pump is assumed to be in contact with an external reservoir with a well defined temperature $T$, for instance  a large quantity of solvent molecules in thermal equilibrium. We consider rates of the form
\begin{eqnarray}
R_{ba}^{(1)} & = & \exp \left[ \frac{E_a - B_1 - \theta f}{T}\right], \nonumber \\
R_{ab}^{(1)} & = & \exp \left[ \frac{E_b - B_1 +\left(1- \theta\right) f}{T}\right], \nonumber \\
R_{ba}^{(2)} & = & \exp \left[ \frac{E_a  -B_2 +\left(1- \theta\right) f}{T}\right], \nonumber \\
R_{ab}^{(2)} & = & \exp \left[ \frac{E_b - B_2 - \theta f}{T}\right].
\label{Eq:defrates}
\end{eqnarray}
Here $R_{ba}^{(1)}$ is the rate of transitions  from site $a$ to site $b$ that are made via mechanism $1$. Similar notation is used for the rest of the rates. $0\le \theta \le 1$ is a load distribution factor that describes the way that the resisting force is divided between two opposite transitions. 

Two comments regarding the rates in Eq. (\ref{Eq:defrates})  are in order. These rates were brought to a simple form that would simplify the subsequent calculations. Specifically, we work in units where $k_B=1$. We also measure energies and barriers with respect to reference values $E_0$ and $B_0$.  We then rescale time to absorb the factor of
$\tilde{\nu}= \nu \exp \left[-\frac{B_0-E_0}{T}\right]$ that is common to all rates. This crucially means that $E>B$ in (\ref{Eq:defrates}) does not mean that the metastable state is destroyed. In fact, we assume that the sites are stable for all the values of the site energies and barriers used below, and that as a result the thermal activation description of the transitions remains valid.

The two-site system can be operated as a stochastic pump by varying the two energies $E_a (t), E_b (t)$, and the two barriers $B_1(t), B_2(t)$
periodically in time. In contrast, the temperature and the resisting force are kept fixed. In the following, we also assume that we are given a period $\tau$, so that the externally controlled parameters
must satisfy $E_i (t+\tau)=E_i (t)$ and $B_j (t+\tau)=B_j (t)$. Note that demanding an external period of $\tau$ allows for time dependence that exhibits smaller periods, as long as an integer number of periods are completed within time $\tau$.

The probability to find the system in the sites, $P_a (t)$ and $P_b (t)=1-P_a (t)$, evolves according to the time dependent master equation
\begin{equation}
\frac{d \vec{P}}{d t} = \mathbf{R}(t) \vec{P} (t),
\label{Eq:master}
\end{equation}
with $ \vec{P}= \left(\begin{smallmatrix} P_a \\ P_b \end{smallmatrix} \right)$, and 
\begin{equation}
\mathbf{R} (t) = \left( \begin{matrix} -R_{ba}^{(1)}- R_{ba}^{(2)} & R_{ab}^{(1)}+R_{ab}^{(2)} \\  R_{ba}^{(1)}+ R_{ba}^{(2)} & -R_{ab}^{(1)}-R_{ab}^{(2)} \end{matrix} \right).
\label{Eq:defR}
\end{equation}
When the transition rates are periodic in time the master equation has a periodic asymptotic solution satisfying $\vec{P}^{(ps)} (t+\tau)=\vec{P}^{(ps)} (t)$. In the following  
we will be interested in properties of the pump in this periodic solution, ignoring transient solutions. We will therefore omit the superscript $(ps)$ in the rest of the paper, since
we will never consider any other solutions of the master equation.

Changing the energies and barriers periodically does not ensure that the system is operated as a stochastic pump. Calling the system a pump implies that the 
external variation of parameters is the source of energy that drives the system, and that this energy is transduced to some useful form of work. Here this means work done against the resisting force $f$. This is achieved by making more
$a \rightarrow b$ than $b \rightarrow a$ transitions using mechanism $1$, and similarly more $b \rightarrow a$ transitions using mechanism $2$.

This qualitative description can be made precise by calculating the energy that the pump exchanges with the outside world. The work that is done on the pump in each cycle
is given by
\begin{equation}
W= \int_t^{t+\tau} d t^\prime \left[ \dot{E}_a (t^\prime) P_a (t^\prime)+ \dot{E}_b (t^\prime) P_b (t^\prime)\right].
\label{Eq:defW}
\end{equation}
The useful output is due to transitions against the resisting force. This definition is motivated by the ability to interpret the transitions as involving the storage or release of useful energy, e.g. by lifting a mass, or compressing a spring.
The energy output during a cycle is given by
\begin{equation}
\Phi =  f \int_t^{t+\tau} d t^\prime \left[ R_{ba}^{(1)} (t^\prime) P_a (t^\prime) -  R_{ab}^{(1)} (t^\prime) P_b (t^\prime)  +R_{ab}^{(2)} (t^\prime) P_b (t^\prime) -  R_{ba}^{(2)} (t^\prime) P_a (t^\prime)  \right].
\label{Eq:defoutput}
\end{equation}
The system is operated as a stochastic pumps when $W \ge \Phi > 0$. In this regime the pump's efficiency is
\begin{equation}
\eta = \frac{\Phi}{W}.
\label{eq:defeff}
\end{equation}

\section{Stating the optimization problem}
\label{sec:limitations}

Our goal is to find the periodic functions $E_a (t), E_b (t), B_1 (t)$ and $B_2 (t)$ that would result in maximal output $\Phi$. 
For a fixed period this is equivalent to maximizing the output power $\Phi/\tau$.
Equation (\ref{Eq:defoutput}) allows us to 
express the output as a functional of the energies and barriers. Some consideration reveals that this optimization problem is not well posed.
During times in the cycle when probability flows from $a$ to $b$ it is beneficial to take $E_a (t) \rightarrow \infty$ and $E_b (t) \rightarrow -\infty$ and also 
$B_2  (t) \rightarrow \infty$ and $B_1 (t) \rightarrow - \infty$. Similar considerations apply when the probability flows from $b$ to $a$.


This tendency of sending the parameters to infinity is unphysical and therefore unwanted. It is not realistic to expect that in any real world realization of a stochastic pump
the experimentalists will have arbitrary freedom to control the system's parameters. We will therefore assume that the energies can be varied
in a limited range
\begin{equation}
0 \le E_a (t), E_b(t) \le E_{max}.
\end{equation}
Similarly, we will demand that both barriers satisfy
\begin{equation}
0 \le B_1 (t), B_2(t).
\end{equation}
Note that we do allow the barriers to be arbitrarily large, since being able to completely prevent transitions at certain times can be beneficial, and is not that different from the solution that would be found when using a finite, but very large  barrier height. We note that changing the lower bound of barriers can be chosen to be $0$, for convenience, since it is measured from an almost arbitrary reference point, as discussed earlier.

With these restrictions on the values of the energies and barriers, the problem of finding the driving protocol that would maximize the output (\ref{Eq:defoutput}) seems well posed. As will become clear in the following, even such a well posed problem can exhibit singular optimal solutions.

\section{A family of candidates for the optimal solution}
\label{sec:cycle}

The model depicted in Fig. \ref{fig:motivation}(c) is simple enough for one to use physical intuition to identify driving protocols that are good candidates for optimal solutions. Consider a driving protocol with a period $\tau$. In any non trivial cycle, the probability must flow between sites $a$ and $b$. As the overall phase of the cycle is arbitrary, we can assume that probability
flows from $a$ to $b$ in the first half of the cycle, and in the opposite direction in the second half.
We can try to maximize the output in the first half of the cycle by preventing transitions through $2$, since they would contribute negatively. Similarly, we gain more output by increasing the energy difference $E_a-E_b$. This results in $E_a (t)=E_{max}$, $E_b (t)=0$, $B_1 (t)=0$, and $B_2 (t)= \infty$ for $0 \le t < \frac{\tau}{2}$.
In the other half of the cycle probability flows from $b$ to $a$. Repeating the same considerations suggests taking
$E_a (t) =0$, $E_b (t)=E_{max}$, $B_1 (t)= \infty$, and $B_2 (t)=0$ for $\frac{\tau}{2} \le t < \tau$. Below we calculate the solution of the master equation with this driving protocol, and study its properties, focusing on the output.

This driving cycle has pieceewise constant rates, which can attain only two non vanishing values. For $0 \le t < \frac{\tau}{2}$
we have 
\begin{eqnarray}
R_{ba}^{(1)} (t)& =& \exp \left[ \frac{E_{max}-f \theta}{T}\right] \equiv K_1, \\
R_{ab}^{(1)}(t) & = & \exp \left[ \frac{f (1-\theta)}{T} \right] \equiv K_2, 
\end{eqnarray}
and $R_{ba}^{(2)} (t)= R_{ab}^{(2)} (t)=0$. The transition rate matrix in this part of the cycle is given by
\begin{equation}
\mathbf{R}_1 = \left( \begin{matrix} -K_1 & K_2 \\ K_1 & - K_2 \end{matrix} \right).
\end{equation}
The propagator for the first half of the cycle can be calculated with the help of the spectral decomposition of $\mathbf{R}_1$, see appendix \ref{sec:decomp} for details.
We find
\begin{equation}
\mathbf{U}_1 = e^{\mathbf{R}_1 \frac{\tau}{2}}=  \frac{1}{K_1+K_2} \left( \begin{matrix} K_2+x K_1 & K_2 (1-x) \\ K_1 (1-x) & K_1+x K_2 \end{matrix} \right),
\label{eq:U1}
\end{equation}
where $x \equiv \exp\left[ -\left( K_1 + K_2 \right) \frac{\tau}{2}\right]$ is a useful abbreviation.

The second half of the cycle is treated in the same way. Here
\begin{equation}
\mathbf{R}_2 = \left( \begin{matrix} -K_2 & K_1 \\ K_2 & - K_1 \end{matrix} \right),
\label{eq:r2}
\end{equation}
and
\begin{equation}
\mathbf{U}_2= e^{\mathbf{R}_2 \frac{\tau}{2}}=  \frac{1}{K_1+K_2} \left( \begin{matrix} K_1+x K_2 & K_1 (1-x) \\ K_2 (1-x) & K_2+x K_1 \end{matrix} \right).
\end{equation}

The periodic solution of the master equation satisfies 
\begin{equation}
\vec{P} (\tau)=\mathbf{U}_2 \mathbf{U}_1 \vec{P} (0).
\label{eq:demandperiod}
\end{equation}
The two-site model exhibits some symmetry. In particular, the states $a$ and $b$ play equivalent roles at different times during the cycle. Here, this means that the dynamics during the second half of the cycle is identical to that of the first half as long as one replaces $a \leftrightarrow b$ and $1 \leftrightarrow 2$.
One can therefore replace Eq. (\ref{eq:demandperiod}) with
\begin{equation}
\left( \begin{matrix} P_b (0) \\ P_a (0)\end{matrix}\right) = \left( \begin{matrix} P_a (\tau/2) \\ P_b (\tau/2)\end{matrix}\right) = \mathbf{U}_1  \left( \begin{matrix} P_a (0) \\ P_b (0)\end{matrix}\right).
\label{eq:halfperiod}
\end{equation}
After a straightforward calculation one finds
\begin{equation}
\label{eq:p0}
P_a (0) = \frac{K_1 + x K_2}{\left(K_1 + K_2 \right) \left( 1+x\right)},
\end{equation}
and
\begin{equation}
\label{eq:p0b}
P_b (0) =1-  P_a (0) = \frac{K_2 + x K_1}{\left(K_1 + K_2 \right) \left( 1+x\right)}.
\end{equation}
The probability after half a cycle can be found with the help of (\ref{eq:halfperiod}), e.g. $P_a (\tau/2)=P_b (0)$.

Calculation of the power output of the pump is simplified considerably for this cycle, because at any given time only transitions that use one of the mechanisms are possible. This creates a strong coupling between the currents and the changes of probabilities during the cycle. The contribution of transitions through link $1$ to the output comes from the first half of the cycle, and is equal to $f \left( P_a (0)-P_a (\tau/2)\right)$. Taking into account also a similar contribution from the second half of the cycle, we have
\begin{equation}
\Phi = f \left( P_a (0)-P_a (\tau/2)\right) + f \left( P_b (\tau/2)-P_b (\tau)\right).
\end{equation}
Substituting the value of the probabilities at different times results in
\begin{equation}
\Phi  = \frac{2f (K_1-K_2 ) (1-x)}{(K_1+K_2 ) (1+x)}.
\label{eq:phisimple}
\end{equation}
Note that this output is indeed positive as long as $K_1 > K_2$, or equivalently $E_{max}>f$.

The work that is done on the pump originates from the changes in the energies at times $t=0,\frac{\tau}{2}$. It is given by
\begin{equation}
W = E_{max} \left[ P_a (0) -P_b (0) - P_a \left( \frac{\tau}{2}\right) + P_b \left( \frac{\tau}{2} \right) \right].
\end{equation}
 Substituting the probabilities leads to
\begin{equation}
W  = \frac{2E_{max} (K_1-K_2 ) (1-x)}{(K_1+K_2 ) (1+x)}.
\end{equation}
The efficiency of the pump in this driving cycle turns out to be particularly simple
\begin{equation}
\eta = \frac{f}{E_{max}}.
\end{equation}
An equivalent expression for the efficiency was previously found in a similar three-site model of a stochastic pump \cite{Rahav2011}.

We point out that the $\tau$-periodicity of energies allows for solutions that are comped of a combination of cycles with smaller periods. Of particular interest are ones composed of $n$ repetitions of a cycle with a period $\tau^\prime=\tau/n$. The solution given above is easily modified to apply for the these cycles, meaning that the calculations above actually describes to a family of solutions with different periods.


\section{Local optimality of the driving protocols}
\label{sec:stab}

In this section we examine the stability of the solutions studied in Sec. \ref{sec:cycle} to small perturbations. Specifically, we examine small changes in the switching time and in the energies and barriers, but not in the overall period $\tau$. We find below that such changes reduce the output of the cycles, meaning that these cycles locally maximize the output. 

\subsection{Stability to small changes in the switching time}

Let us consider a cycle that is deformed by slightly modifying the duration of the two half cycles, so that the overall period of the cycle is kept fixed. Without losing generality, we can assume
that the time in which the rate matrix is changes from $\mathbf{R}_1$ to $\mathbf{R}_2$ is now $\tau/2 +\delta t$ instead of $\tau/2$. We wish to find out how such a change will affect the output of the cycle. For small $\delta t$ this can be studied using perturbation theory. The calculation is almost straightforward, but care should be taken, because the perturbation also results in small corrections to the periodic solution of the master equation (\ref{Eq:master}), and they must be taken into account.

The propagator of the first segment of the cycle is now given by
\begin{equation}
\mathbf{U}^\prime_1= e^{\mathbf{R}_1 \left( \frac{\tau}{2}+\delta t\right)} = \mathbf{U}_1 + \delta t x \mathbf{R}_1  -\frac{\delta t^2 x}{2} \left( K_1+K_2\right) \mathbf{R}_1 + O \left( \delta t^3 \right).
\end{equation}
Here we kept terms up to order $\delta t^2$, since it will become clear that this is the leading order correction to the output of the cycle. Similarly, the propagator of the second segment of the cycle is
\begin{equation}
\mathbf{U}^\prime_2= e^{\mathbf{R}_2 \left( \frac{\tau}{2}-\delta t\right)} = \mathbf{U}_2 - \delta t x \mathbf{R}_2  -\frac{\delta t^2 x}{2} \left( K_1+K_2\right) \mathbf{R}_2 + O \left( \delta t^3 \right).
\end{equation}
The propagator of the full cycle is simply the product $\mathbf{U}^\prime_2 \mathbf{U}^\prime_1 $. To second order we find
\begin{equation}
\mathbf{U}^\prime  =  \mathbf{U}_2 \mathbf{U}_1 + \delta t x \left( K_1 - K_2\right)  \left( \begin{matrix}-1 & -1 \\ 1&1 \end{matrix}\right)+\frac{\delta t^2}{2} x \left( K_1^2 -K_2^2\right)   \left( \begin{matrix}-1 & -1 \\ 1&1 \end{matrix}\right) + O \left( \delta t^3 \right).
\label{eq:propprime}
\end{equation}
The appearance of the matrix $\left( \begin{smallmatrix}-1 & -1 \\ 1&1 \end{smallmatrix}\right) $ is a result of conservation of probability and the fact that the system has only two sites. As a result, 
deviations from the unperturbed periodic solutions are always proportional to $\left( \begin{smallmatrix} 1 \\ -1 \end{smallmatrix} \right)$ to avoid violating the demand that $P_A^\prime (t)+P_B^\prime (t)=1$. Multiplication of the matrix above and any normalized probability distribution results in a term which is proportional to the vector $\left( \begin{smallmatrix} 1 \\ -1 \end{smallmatrix} \right)$. 

The periodic solution of the perturbed cycle can be found with the help of Eq. (\ref{eq:demandperiod}), where one uses the modified propagator (\ref{eq:propprime}).
Some algebra results in
\begin{equation}
P_A^\prime (0) = \frac{K_1+x K_2}{\left( K_1+K_2\right) (1+x)} - \delta t \frac{x}{1-x^2} (K_1 - K_2) -\frac{\delta t^2}{2} \frac{x}{1-x^2} \left( K_1^2-K_2^2\right) + O \left( \delta t^3 \right),
\end{equation}
where, as always, $P_B^\prime (0) =1-P_A^\prime (0) $. The calculation of the output requires knowledge of the probabilities at the switching time $t=\tau/2+\delta t$. These can be calculated from the probabilities at $t=0$ with the help of the propagator $\mathbf{U}^\prime_1$. A short calculation results in
\begin{equation}
P_A^\prime \left( \frac{\tau}{2}+ \delta t \right) = \frac{K_2+x K_1}{\left( K_1+K_2\right) (1+x)} - \delta t \frac{x}{1-x^2} (K_1 - K_2) +\frac{\delta t^2}{2} \frac{x}{1-x^2} \left( K_1^2-K_2^2\right) + O \left( \delta t^3 \right).
\end{equation}

The calculation of the output is essentially the same as for the unperturbed cycle. During the first segment of the cycle a probability of $P_A^\prime (0) - P_A^\prime \left( \frac{\tau}{2}+\delta t \right)$ flows from
site $a$ to site $b$. In the second segment of the cycle the same probability flows back. In both segments one of the pathways is open and the other is closed, so that this flow must contribute positively to the output. As a result the output is given by
\begin{equation}
\Phi^\prime = 2 f \left [P_A^\prime (0) - P_A^\prime \left( \frac{\tau}{2}+\delta t \right)\right] =\frac{2 f\left( K_1-K_2\right)(1-x)}{\left( K_1+K_2\right) (1+x)}-\delta t^2 \frac{2 f x}{1-x^2} \left( K_1^2-K_2^2\right) + O \left( \delta t^3 \right).
\label{eq:outputpertt}
\end{equation}
The first term on the right hand side of  Eq. (\ref{eq:outputpertt}) is the output of the unperturbed cycle.
We note that $0<x<1$ and $K_1>K_2$. It is therefore clear that small changes in the switching time always reduce the output of a cycle, as long at overall period is unchanged. The argument also covers changes in the time of the switch from $\mathbf{R}_2$ to $\mathbf{R}_1$. This can be seen by changing the origin of time to the time when 
$\mathbf{R}_2$ is replaced by $\mathbf{R}_1$, which maps the problem to the one studied above.

\subsection{Linear response corrections due to small changes in the barriers and energies}

We now consider how the power output of the solutions discussed in Sec. \ref{sec:cycle} is affected if the time dependence of the energies and barriers is slightly modified. To this end we write
\begin{equation}
\mathbf{R}^\prime (t) = \mathbf{R} (t) + \delta \mathbf{R} (t),
\end{equation}
where both $\mathbf{R} (t) $ and $ \delta \mathbf{R} (t)$ are periodic with the same period $\tau$.
In the following, we consider $\delta \mathbf{R} (t)$ to be small. We then examine the linear order corrections to the power output from such perturbations.
The small changes in the rates result in a perturbed periodic state
\begin{equation}
\vec{P}^\prime (t) = \vec{P} (t) + \delta \vec{P} (t).
\end{equation}
Here $\vec{P}$ is the solution of the unperturbed problem
\begin{equation}
\frac{d \vec{P}}{dt}= \mathbf{R} (t) \vec{P} (t).
\end{equation}
This unperturbed solution can be calculated explicitly using the spectral decomposition found in Appendix \ref{sec:decomp} and the initial condition (\ref{eq:p0}).
A straightforward calculation results in
\begin{equation}
\vec{P} (\tilde{t}) = \frac{1}{\left( K_1+K_2\right)(1+x)} \left\{ \begin{matrix} \left( \begin{matrix} K_2 (1+x) + x_{\tilde{t}} \left( K_1-K_2 \right) \\ K_1 (1+x) + x_{\tilde{t}} \left( K_2-K_1 \right) \end{matrix}\right)& \;\;\;\; 0 \le \tilde{t} <\frac{\tau}{2}\\  \left( \begin{matrix} K_1 (1+x) + x_{\tilde{t}-\tau/2 } \left( K_2-K_1 \right) \\ K_2  (1+x) + x_{\tilde{t}-\tau/2 } \left( K_1-K_2 \right) \end{matrix}\right) & \;\;\;\;\frac{\tau}{2} \le \tilde{t} < \tau . \end{matrix} \right.
\end{equation}
Here $\tilde{t}\equiv t \mod \tau$, and $x_t \equiv \exp \left[ - (K_1+K_2 ) t\right]$. We use this compact notation since exponentials like  $x_t$  are ubiquitous in the following calculations.
We remind the reader that  $x_{\tau/2}=x$ as this specific term appears quite often. It is easy to verify that this solution is consistent with Eqs. (\ref{eq:p0}) and (\ref{eq:p0b}).

The linear correction to the power output is obtained by taking the leading order variation of Eq. (\ref{Eq:defoutput}). We find
\begin{equation}
\delta \Phi = f \int_0^\tau d t^\prime \left[ \delta R_{ba}^{(1)} P_a - \delta R_{ab}^{(1)} P_b + \delta R_{ab}^{(2)} P_b - \delta R_{ba}^{(2)} P_a +
R_{ba}^{(1)} \delta P_a - R_{ab}^{(1)} \delta P_b + R_{ab}^{(2)} \delta P_b - R_{ba}^{(2)} \delta P_a \right].
\label{eq:tempdeltap}
\end{equation}
To be able to say something useful  about the sign of $\delta \Phi$ we need  to rewrite this quantity as a time integral over the change of the rates, $\delta R_{ij}(t)$, multiplied by known functions of time. The first four terms in Eq. (\ref{eq:tempdeltap}) already have the desired form, but the last four terms do not. To proceed, we first find explicit expressions for the linear corrections to the periodic state. The details are given in Appendix \ref{sec:dp}. 

The symmetry of the model and cycle can be used to simplify the calculations. As mentioned above, in the unperturbed solution the two sites and barriers play the same role at different parts of the cycle.  As a result a change in $E_a$ at time $0<t<\frac{\tau}{2}$ plays the same role as the same change in $E_b$ at time $t+\frac{\tau}{2}$.  This means any perturbation in the second half of the cycle
has an equivalent counterpart in the first half. Moreover, in linear order the contributions from different times are additive. It is therefore enough to show that $\delta \Phi$ is negative to all possible small changes of rates in the first half of the cycle. Substituting the value of the unperturbed rates, we find
\begin{equation}
\delta \Phi = f \int_0^\frac{\tau}{2} d t^\prime \left[ \delta R_{ba}^{(1)} P_a - \delta R_{ab}^{(1)} P_b + \delta R_{ab}^{(2)} P_b - \delta R_{ba}^{(2)} P_a \right]
+ f \left( K_1 +K_2 \right) \left[ \int_0^\frac{\tau}{2}d t^\prime \delta P_a (t^\prime) -\int_\frac{\tau}{2}^\tau dt^\prime \delta P_a (t^\prime)\right].
\end{equation}
 
By integrating Eq. (\ref{eq:usfulpa}) from $0$ to $\frac{\tau}{2}$, and from $\frac{\tau}{2}$ to $\tau$, one can obtain
\begin{multline}
 \left( K_1 +K_2 \right) \left[ \int_0^\frac{\tau}{2}d t^\prime \delta P_a (t^\prime) -\int_\frac{\tau}{2}^\tau dt^\prime \delta P_a (t^\prime)\right] = 2 \left[ \delta P_a (0) -\delta P_a \left( \frac{\tau}{2}\right)\right] \\ + \int_0^\frac{\tau}{2} d t^\prime \left[ \left( \delta R_{ab}^{(1)}+\delta R_{ab}^{(2)}\right) P_b - \left( \delta R_{ba}^{(1)}+\delta R_{ba}^{(2)}\right) P_a \right].
\end{multline}
With the help of Eq. (\ref{eq:dpdiff}) the linear correction for the output can be written as
\begin{equation}
\delta \Phi = 2 f \int_0^\frac{\tau}{2} d t^\prime \left\{ \left[1-\frac{1}{1+x} e^{-(K_1+K_2)\left[\frac{\tau}{2}-t^\prime \right]} \right] \left( \delta R_{ab}^{(2)} P_b -   \delta R_{ba}^{(2)} P_a \right) + \frac{1}{1+x}   e^{-(K_1+K_2)\left[\frac{\tau}{2}-t^\prime \right]} \left( \delta R_{ba}^{(1)} P_a -  \delta R_{ab}^{(1)} P_b\right) \right\}.
\label{eq:middeltaphi}
\end{equation}

The possible perturbations of the rates are not independent, since for instance, a small change in $E_a$ will affect both $\delta R_{ba}^{(1)}$ and $\delta R_{ba}^{(2)}$. We wish to write the linear correction to the output as a sum over independent terms. It will be convenient to work with the variables ${\cal E}_{a,b} = e^{E_{a,b}/T}$ and ${\cal B}_{1,2}=e^{-B_{1,2}/T}$. The rates can be expressed in terms of these variables as $R_{ba}^{(1)}={\cal E}_a {\cal B}_1 e^{-f \theta /T}$, $R_{ab}^{(1)}= {\cal E}_b {\cal B}_1 e^{f (1-\theta)/T}$ , $R_{ba}^{(2)}= {\cal E}_a {\cal B}_2 e^{f(1-\theta)/T}$ and $R_{ab}^{(2)}= {\cal E}_b {\cal B}_1 e^{-f\theta/T}$. 

We now focus on perturbations made in the first half of the cycle.  $0\le t <\frac{\tau}{2}$. During the first half of the cycle ${\cal E}_a (t)= {\cal E}_{max}=e^{E_{max}/T}$ and ${\cal E}_b (t) =1$, and only perturbations with $\delta {\cal E}_a (t) \le 0$ and $\delta {\cal E}_b (t) \ge 0$ are admissible. For the barriers we have ${\cal B}_1 (t)=1$ and ${\cal B}_2 (t)=0$ and as a result the perturbations must have $\delta {\cal B}_1 (t) \le 0$ and $\delta {\cal B}_2 (t)\ge 0$. The linear variation of the rates during the first half of the cycle can the be recast as
\begin{eqnarray}
\delta R_{ba}^{(1)} & = & \delta {\cal E}_a e^{-f \theta /T} +\delta {\cal B}_1 {\cal E}_{max} e^{- f \theta/T} = \delta {\cal E}_a K_1/{\cal E}_{max} +\delta {\cal B}_1 K_1,  \nonumber \\
\delta R_{ab}^{(1)} & = & \delta {\cal E}_b e^{f(1-\theta)/T} + \delta {\cal B}_1 e^{f(1-\theta)/T} =  \delta {\cal E}_b K_2 +  \delta {\cal B}_1 K_2, \nonumber \\
\delta R_{ba}^{(2)} & = & \delta {\cal B}_2 K_1 e^{f/T} =  \delta {\cal B}_2  {\cal E}_{max} K_2, \nonumber \\
\delta R_{ab}^{(2)} & = & \delta {\cal B}_2 K_1/{\cal E}_{max}.
\label{eq:pertrates}
\end{eqnarray}

Substitution of (\ref{eq:pertrates}) allows us to rewrite Eq. (\ref{eq:middeltaphi}) in the form
\begin{equation}
\delta \Phi = 2 f \int_0^{\frac{\tau}{2}} d t^\prime \left[ \delta {\cal E}_a (t^\prime) g_a (t^\prime) +  \delta {\cal E}_b (t^\prime) g_b (t^\prime) +  \delta {\cal B}_1 (t^\prime) g_1 (t^\prime) + \delta {\cal B}_2 (t^\prime) g_2 (t^\prime) \right].
\end{equation}
We find that $g_a(t) = \frac{K_1}{(1+x) {\cal E}_{max} } e^{-(K_1+K_2)\left[\frac{\tau}{2}-t \right]} P_a (t) \ge 0$. Since $\delta {\cal E}_a (t) \le 0$, one sees that small changes in $E_a$ must result in a smaller output. Noting that $g_b (t) = - \frac{K_2}{(1+x)} e^{-(K_1+K_2)\left[\frac{\tau}{2}-t \right]} P_b  (t) \le 0$ and that $\delta {\cal E}_b \ge 0$, we conclude that the same 
applies for small changes in $E_b (t)$.

The dependence on changes in the barriers is somewhat more complicated. One finds $g_1 (t) =  \frac{1}{1+x}   e^{-(K_1+K_2)\left[\frac{\tau}{2}-t^\prime \right]} \\ \times \left\{ K_1 P_a  (t) -  K_2 P_b (t)\right\}$. The term in curly brackets is minimal at $t=\frac{\tau}{2}$, where its value is $(K_1+K_2)x/(1+x)$. In fact, $K_1 P_a  (t) -  K_2 P_b (t)$ will reach zero if the system is left to relax to equilibrium with time independent rates. As a result,   $K_1 P_a (t) -  K_2 P_b (t) \ge 0$ for all $0 \le t \le \tau/2$ in the unperturbed solution. We therefore find that $g_1 (t) >0$. Combined with
$\delta {\cal B}_1 (t) \le 0$, we find that reducing the height of the barrier $B_1$ results in a smaller output.

The remaining term involves the function $g_2(t)=  \left[1-\frac{1}{1+x} e^{-(K_1+K_2)\left[\frac{\tau}{2}-t \right]} \right] \left( \frac{K_1}{{\cal E}_{max}}   P_b (t) -{K_2}{{\cal E}_{max}}  P_a  (t) \right) $. The term in the first square brackets is clearly positive. The second term is  largest at $t=\frac{\tau}{2}$, where its value is 
\begin{equation}
 \frac{K_1}{{\cal E}_{max}}   P_b  \left(\frac{\tau}{2} \right)-{K_2}{{\cal E}_{max}}  P_a   \left( \frac{\tau}{2}\right) = \frac{1}{(K_1+K_2)(1+x) {\cal E}_{max}} \left[ K_1^2 \left( 1-e^{\frac{2f}{T}}\right) +x K_1 K_2 \left(1-{\cal E}_{max}^2 \right)\right].
\end{equation}
Noting that $f>0$ and ${\cal E}_{max}>1$ we find that $g_2(t)<0$ for $0\le t < \frac{\tau}{2}$. The allowed perturbations $\delta {\cal B}_2 (t) \ge 0$ can therefore only decrease the output. We just checked that all admissible small changes to the rates always decrease the output.

The conclusion from the preceding calculations is that the cycles studied in Sec. \ref{sec:cycle} locally maximize the output with respect to small changes in the driving cycle. This should not be taken as a rigorous proof that such cycles are indeed the optimal solution. The reason is that we have not shown that that the cycle is maximal also with respect to so-called needle perturbations, which allow for large changes of parameters for an infinitesimal time. Such perturbations are commonly studied in optimal control \cite{LiberzonBook}.  In the next section, a numerical approach will suggest that this solution is unstable to changes in the period $\tau$, which can be interpreted as a needle perturbation. Nevertheless,
the same numerical results show that the optimal cycle can be understood in terms of cycles of the type studied here, by comparing cycles with different periods.

\section{Numerical maximization using a genetic algorithm}
\label{sec:num}

 We complement the analytical arguments given above with a numerical
approach that searches for the optimal solution in an completely different way. Finding out that two independent approaches lead to the same results will greatly strengthens our confidence in the validity of the results.

\subsection{A genetic algorithm}
\label{subsection:genetic}

Numerically searching for the driving protocol that maximizes the output is challenging  due to the need to explore the space of four periodic functions, $E_{a,b} (t)$ and $B_{1,2} (t)$. A gradient based search is likely to wander around in nearly-flat regions of the space, and then be trapped in some local maximum. Instead, we used an
approach that is based on a genetic algorithm \cite{genbook}. The term genetic algorithm is used here to describe a heuristic method of searching for a maximum that is inspired by the way that
populations of living  organisms evolve to improve their fitness.
In evolution, a phenotype relevant for survival can be present in a generation of a population. The phenotype itself results from an expression of genes, and genetic variation
results in variability that allows for natural selection.
The survivors constitute the next generation. This population includes members with random minor changes due to mutations. The process repeats itself and the fitness of the entire population improves, because although most individual changes are detrimental, beneficial changes are more likely to be propagated to the following generations.
Our genetic algorithm operates similarly, with the output as the phenotype, and the time dependence of the energies and barriers as the genetic code.

Specifically, the genetic code of a member of the population is defined as the values of $E_a (t), E_b (t), B_1 (t)$, and $B_2 (t)$ during the cycle. In practice, the cycle was divided into $N$ time segments, $\left[ \tau i /N, \tau(i+1)/N \right)$, with $i=0,1,2,\cdots , N-1$. We used $\tau=2 \pi/6$, and $N=128$ in all the results presented below. In each time segment, the energies and barriers were taken to be constant during the segment. The driving protocol is then identified by recording the energies and barriers in each of the segments, $E_{a,b}(t_i)$ and $B_{1,2} (t_i)$. These  512 values define the genetic code of each driving cycle. For practical reasons, we enforced a maximal value of $B_{max}=10$ to the various barriers.

The selection criterion is, naturally, based on the maximization the output $\phi$ of the system, as defined in section \ref{sec:model}. 
It can be calculated efficiently as described in the following.
 In each of the time segments one can generalize the spectral decomposition presented 
in App. \ref{sec:decomp}  to a rate matrix with arbitrary, but time independent, rates. The resulting propagator for the $i$'th time step is
\begin{equation}
{\mathbf U}_i = \frac{1}{\left| \lambda_2^{(i-1)}\right|} \left( \begin{matrix}R_{ab}^{(i-1)} + e^{\lambda_2^{(i-1)} \left( t_i-t_{i-1}\right)} R_{ba}^{(i-1)} & R_{ab}^{(i-1)} \left[ 1-e^{\lambda_2^{(i-1)} \left( t_i-t_{i-1}\right)}  \right] \\  R_{ba}^{(i-1)} \left[ 1-e^{\lambda_2^{(i-1)} \left( t_i-t_{i-1}\right)}  \right] &  R_{ba}^{(i-1)} + e^{\lambda_2^{(i-1)} \left( t_i-t_{i-1}\right)} R_{ab}^{(i-1)} \end{matrix}\right).
\end{equation}
Here $R^{(i)}_{ab} \equiv R_{ab}^{(1)} (t_i) +  R_{ab}^{(2)} (t_i)$, $R^{(i)}_{ba} \equiv R_{ba}^{(1)} (t_i) +  R_{ba}^{(2)} (t_i)$, and $\lambda_2^{(i)}=-R^{(i)}_{ab}-R^{(i)}_{ba}$.
The propagator of a complete cycle, ${\mathbf U}=\prod_i {\mathbf U}_i$, can then be used to find the periodic solution satisying $\vec{P} (\tau) = {\mathbf U} \vec{P} (0) = \vec{P} (0).$

Once the periodic solution is found, the output of the cycle is calculated by adding contributions from each time step. In the $i$'th time step, ${\mathbf U}_i$ can be used 
to propagte the probabilities $P_{a,b} (t_{i-1})$ to $P_{a,b} (t_{i})$. Then the contibution of this time step to the output of the stochastic pump is calculated using
\begin{multline}
\delta \phi_i = \frac{2f}{\left|\lambda_2^{(i-1)} \right|} \left( t_i-t_{i-1} \right) \left[  R_{ba}^{(1)} (t_{i-1}) R_{ab}^{(2)} (t_{i-1}) - R_{ba}^{(2)} (t_{i-1})   R_{ab}^{(1)} (t_{i-1})\right] \\ + \frac{f}{\left|\lambda_2^{(i-1)} \right|} \left[ R_{ba}^{(1)} (t_{i-1}) + R_{ab}^{(1)} (t_{i-1}) - R_{ba}^{(2)} (t_{i-1}) - R_{ab}^{(2)} (t_{i-1})\right] \left\{P_b (t_i) -P_b (t_{i-1}) \right\}.
\label{eq:outputstep}
\end{multline}
A derivation of Eq. (\ref{eq:outputstep}) can be found in App. \ref{sec:derivationstep}.
Summation of the contributions from all the time steps results in the output of the whole cycle. Numerical calculations that are based on piecewise constant parameters turn out to be 
fast and reliable, since there is no need for naive numerical integration that requires subdivision of the time segments to smaller time steps.

A genetic algorithm that is built upon those calculations is then applied. A population consisting of 200 different driving protocols is generated, and constitutes the first generation. Initially, the energies where drawn randomly while the barriers were set to 1. This population is then subjected to an evolution is which new 'generations' replace old ones by a process of mutations and selection. 
Our algorithm used several types of mutations and other genetic variations to generate new driving protocols:
\begin{enumerate}
\item In small mutations, one of the parameters, i.e. $B_2 (t_i)$, was chosen at random. This value was then modified by adding a random number taken from the range $\pm 2.5 E_{max}/100$. The new value was truncated if it exceeded the allowed range for this parameter.
\item Large mutations are similar, but the change in the parameter could be as large as $\pm 0.5 E_{max}$.
\item Group mutations are similar to small mutations, but the change is applied to all the values of the same parameter in a small time segment (1-5 time steps).
\item A replication is made by picking a time $t_i$ at random, and then copying the values of $E_{a,b} (t_i)$ and $B_{1,2} (t_i)$ onto their counterparts at time $t_{i+1}.$
\item An inversion mutation is created by picking a time $t_i$ and a time segment at random. The segment can have any length between one time step to half of the cycle. For all the parameters in this time segment, one applies the inversion $E \rightarrow E_{max}-E$ and similarly $B \rightarrow B_{max}-B$.
\item Barrier lowering mutations were used to help some systems escape local maximums. An initial time and segment length (between $\tau/128$ and $\tau/2$) were chosen at random. Then one of the energies (namely $a$ or $b$) and one of the barriers ($1$ or $2$) were chosen randomly. The values of the barrier (e.g. $B_2 (t_i)$) in all times during the segment were set to $E_{max}/10$, whereas values of the energy in the segment were set to a time-independent random value (between $0$ and $E_{max}$).
\item In a recombination, two new cycles are created from an existing pair of cycles, by exchanging the values of the parameters between two equal length segments. Both the initial times as well as the length of the segment (between one step and $\tau/2$) were chosen randomly.
\end{enumerate}

A new generation was created from an existing one using the following scheme. The $10$ members with the highest output were passed to the new generation without any modifications. 20 members were created by applying small mutations to these $10$ members. Similarly, 10 members were the result of group mutations. 20 members of the new generations were created by choosing at random a member of the current generation and applying a large mutation. Another 10 members were created using a similar process with replication, 10 member underwent barrier lowering, and further 20 had inversion mutations. Finally, the remaining 100 members were generated
by randomly choosing $50$ pairs of members of the current generation and using recombination. The resulting new generation thus includes a population of 200, the same as its predecessor. This process 
leads to generations that are built from small variations to the members with the highest output, and a variety of members with larger changes to their genetic code, thereby maintaining a diverse population.
The algorithm  progresses between successive generations until it seems that the maximal output observed in each generation saturates. One then hopes that the cycle with the maximal output is indeed the globally optimal solution.

\subsection{Results}

\begin{figure}[htb]
\centering
\subfloat[][]{\includegraphics[width=0.4\textwidth,height=5.1cm]{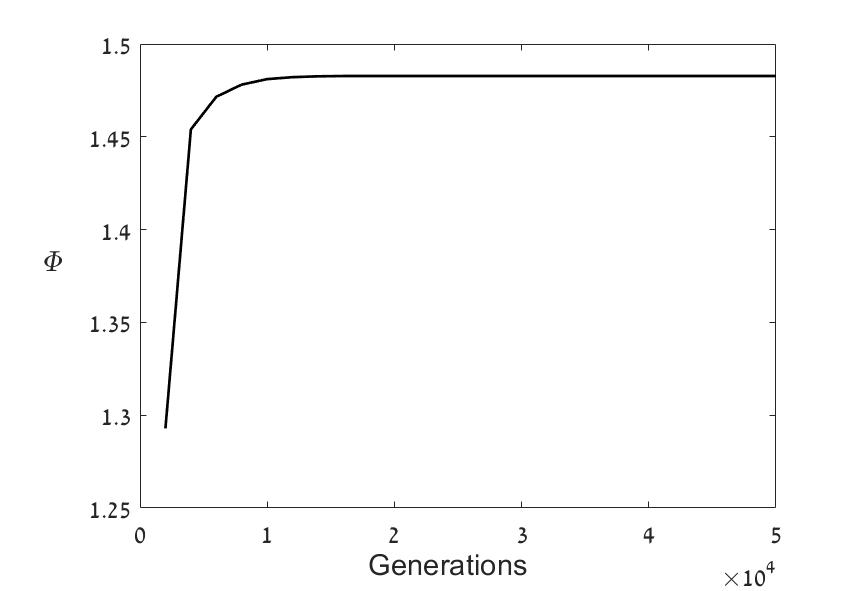}
\label{fig:gene0-a}}
\hspace{8pt}
\subfloat[][]{\includegraphics[width=0.4\textwidth,height=5.1cm]{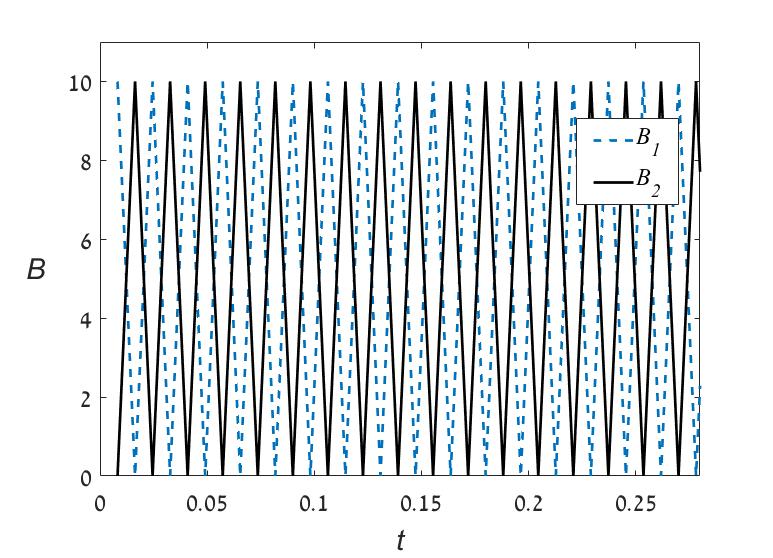}
\label{fig:gene0-b}} \\
\subfloat[][]{\includegraphics[width=0.4\textwidth,height=5.1cm]{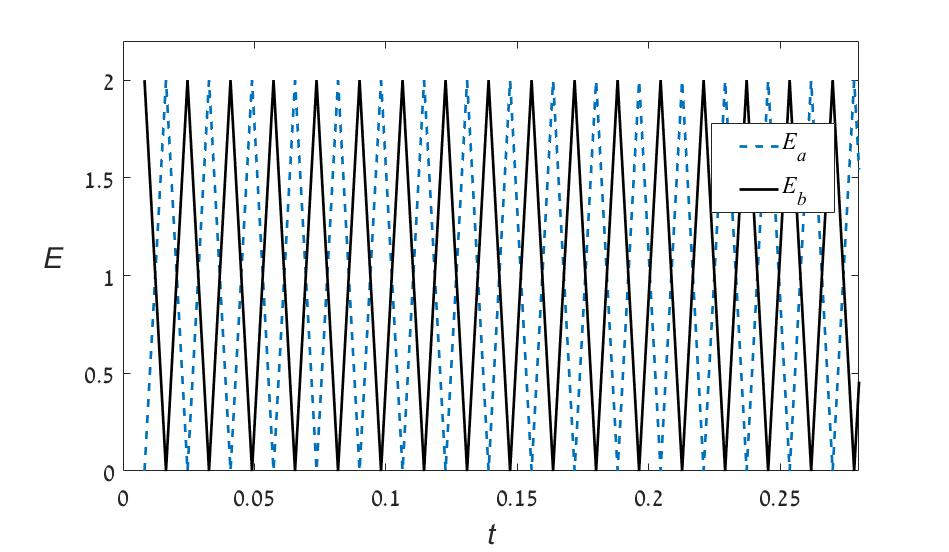}
\label{fig:gene0-c}}
\hspace{8pt}
\subfloat[][]{\includegraphics[width=0.4\textwidth,height=5.1cm]{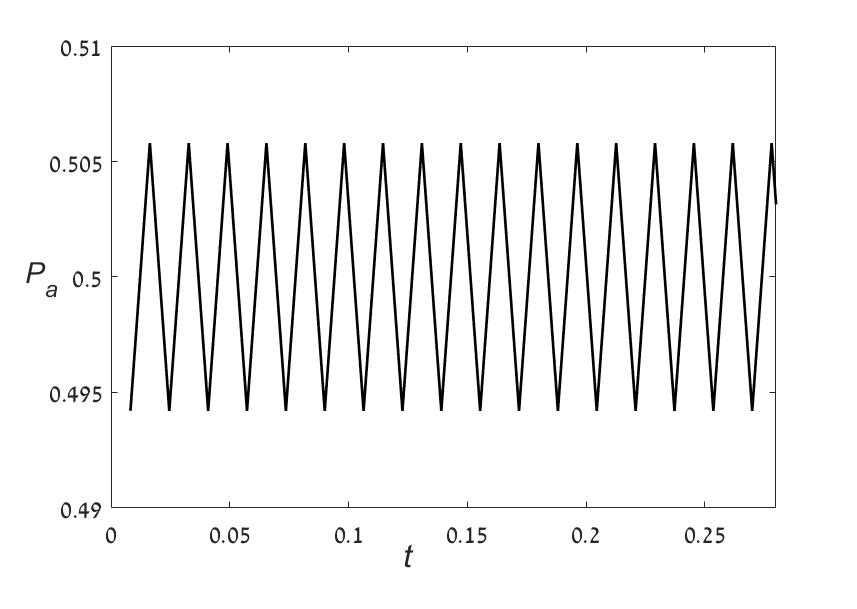}
\label{fig:gene0-d}}
\caption[]{Results of the genetic algorithm:
\subref{fig:gene0-a} Output as a function of the number of generations;
\subref{fig:gene0-b} The time dependence of barriers in the cycle with highest output;
\subref{fig:gene0-c} Time dependence of the site energies in this cycle; and,
\subref{fig:gene0-d} The time dependence of the probability, $P_a (t)$. To avoid clutter only the first quarter or so of the external period $\tau$ are shown in panels \subref{fig:gene0-b}-\subref{fig:gene0-d}. }%
\label{fig:gene0}%
\end{figure}

Fig. \ref{fig:gene0}(a) depicts the variation of the maximal output with different generations. One can see that for a while the output tends to increase with progressing generations. We note that the figure is smoothed by only including points taken every 2000 generations. After a certain number of generations the maximal output
saturates, suggesting convergence to either a deep local or the global maximal output cycle.

Figs. \ref{fig:gene0}(b) and (c) show the time dependence of the energies and barriers of the optimal cycle (according to the genetic algorithm). For clarity only the first quarter of the cycle is depicted, since it is hard to distinguish different lines when the full period is shown. At each time step the value of the parameters match one of the two sets of parameters that were identified in the cycles studied in Sec. \ref{sec:cycle}. To be precise, we find either $E_a=2, E_b=0, B_1=0, B_2=10$ or 
$E_a=0,E_b=2,B_1=10, B_2=0$.  Crucially, the solution that the algorithm found alternates between the two sets of parameters {\it at each time step}. Fig. \ref{fig:gene0}(d) depicts the time evolution of the probability $P_a (t)$
in the optimal cycle. The rapid switching of parameters keeps this probability fairly close to its symmetric value $P_a= P_b=1/2$.

\subsection{Comparison with the analytical approach}

At first, the numerical candidate for an optimal solution may seem at odds with the locally optimal solutions studied in Secs. \ref{sec:cycle} and \ref{sec:stab}. However, a careful consideration reveals that our discussion there assumed that the period $\tau$ of the cycle was fixed, whereas we actually demanded that $E_a(t)=E_a(t+\tau)$, etc. This is consistent with cycles with a period of $\tau/n$ for any
positive integer $n$, as well as other, suboptimal combinations of cycles with smaller periods. For cycles with period $\tau/n$  the  output accumulated after a time $\tau$ is
\begin{equation}
\Phi_n (\tau)  = n \frac{2 f \left(K_1-K_2 \right)}{K_1+K_2} \frac{1-\exp \left[ -(K_1+K_2) \frac{\tau}{2n}\right]}{1+\exp \left[ -(K_1+K_2) \frac{\tau}{2n}\right]}.
\label{eq:phin}
\end{equation}
$\Phi_n (\tau)  $ increases with $n$, meaning that the maximal output is obtained in the limit of vanishing cycle period. The output accumulated after a time $\tau$ is therefore bounded by
\begin{equation}
\Phi^* = \lim_{n \rightarrow \infty} \Phi_n (\tau)  = \frac{f \tau}{2} \left( K_1-K_2\right).
\label{eq:maxpower}
\end{equation}
This observation fits well with the numerical results. The numerical algorithm converged to the solution with the smallest cycle period that was available to it. It suggests that the two-site stochastic pump is operated at maximum output when a singular driving protocol is used. In this singular cycle the pump alternates between the two sets of parameters at an infinite rate,
meaning that one can not assign a definite value for $E_{a,b} (t)$ and $B_{1,2} (t)$.
The resulting maximal power output is given by
\begin{equation}
\dot{\Phi}^* = \frac{f}{2} e^{-\frac{f \theta}{T}} \left[ e^{\frac{E_{max}}{T}}- e^{\frac{f}{T}}\right],
\end{equation}
which can be used to numerically calculate the resisting force $f$ that would increase the power further. In the linear response regime one finds $f^*=\frac{E_{max}}{2}$, as expected.

Solutions that involve infinitely fast switching of the external driving are known in the field of control theory \cite{ChatteringC,SlidingC}.
The terms chattering and sliding control are used to describe various control solutions with this property. The combination of numerical results and the analytical arguments given here suggests that
a solution exhibiting such rapid switching is the optimal solution for the problem studied here. 

\section{Comparison of regularized analytical and numerical optimal solutions}
\label{sec:eps}

A possible problem with the results presented in the previous section is that the numerically obtained optimal solution was restricted by the discrete times steps we
used. One may be worried that the resulting driving cycle is a numerical artifact rather than the best possible approximation for the unconstrained optimal solution. It is therefore desirable to test the genetic algorithm in situations where its convergence is less affected by this specific externally placed restriction.
In this section we chose to modify the criterion for optimization to penalize too frequent parameter switching events. The modified optimal solutions are regularized to exhibit finite periods.

In the genetic algorithm we apply this scheme by defining a cost that includes a (typically small) penalty to changes in site energies
\begin{equation}
C \left( \epsilon, \tau\right)\equiv \Phi (\tau) - \epsilon \sum_{i=1}^N \sum_{\alpha=a,b} \left|E_{\alpha} (t_{i+1})-E_{\alpha (t_i}) \right|.
\label{eq:defcost}
\end{equation}
Here $\Phi (\tau)$ is the accumulated output up to time $\tau$. We then use the genetic algorithm to look for the solutions that maximize this cost.
For small values of $\epsilon$ we expect that this penalty will not drastically change the nature of the solutions, in the sense that they still switch between the same two sets of variables, albeit at a finite rate.

\begin{figure}[htb]
\centering
\subfloat[][]{\includegraphics[width=0.4\textwidth,height=5.1cm]{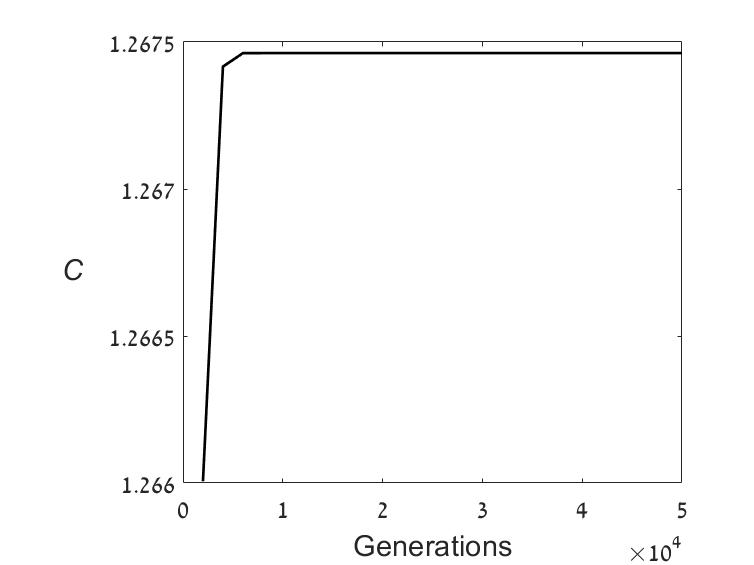}
\label{fig:gene44-a}}
\hspace{8pt}
\subfloat[][]{\includegraphics[width=0.4\textwidth,height=5.1cm]{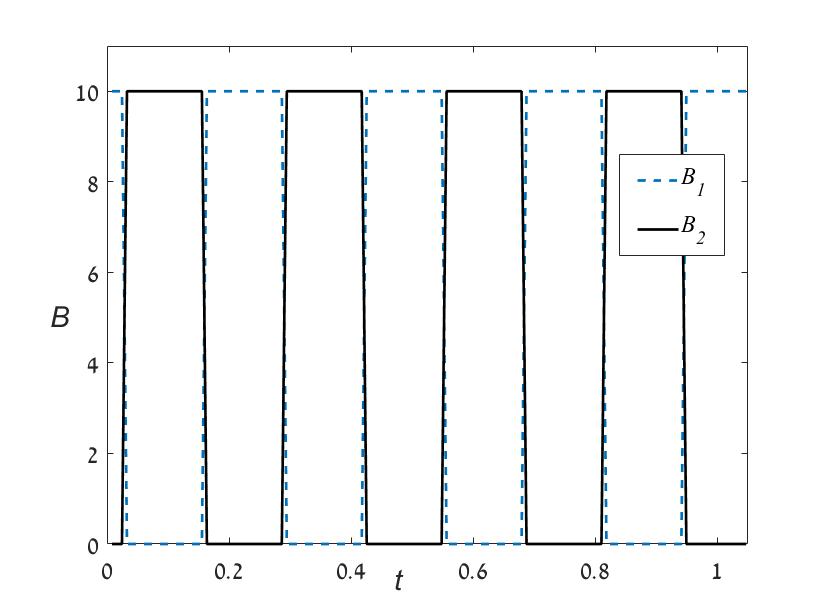}
\label{fig:gene44-b}} \\
\subfloat[][]{\includegraphics[width=0.4\textwidth,height=5.1cm]{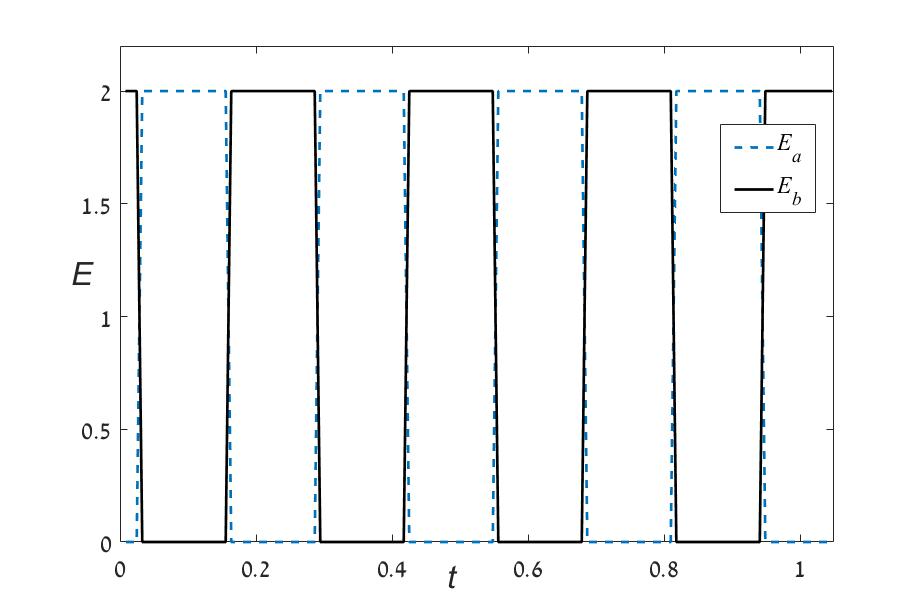}
\label{fig:gene44-c}}
\hspace{8pt}
\subfloat[][]{\includegraphics[width=0.4\textwidth,height=5.1cm]{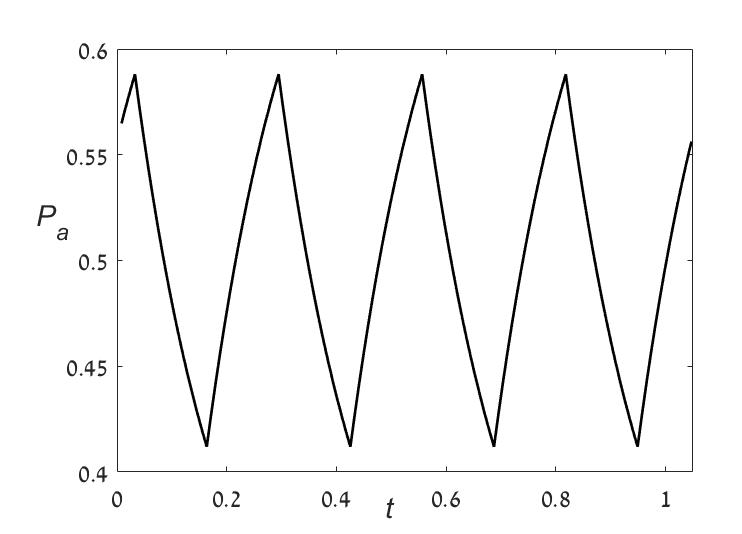}
\label{fig:gene44-d}}
\caption[]{Results of the genetic algorithm for optimization with respect to the cost (\ref{eq:defcost}):
\subref{fig:gene44-a} The cost as a function of the number of generations;
\subref{fig:gene44-b} The time dependence of barriers in the converged solution;
\subref{fig:gene44-c} Time dependence of the site energies in the solution; and,
\subref{fig:gene44-d} The time dependence of the probability to reside in site $a$. }%
\label{fig:gene44}%
\end{figure}

Fig. \ref{fig:gene44} depicts the results of the algorithm for $\epsilon=0.0044$.
Fig. \ref{fig:gene44}(a) shows the maximal value of the cost for different generations. The curve is smoothed by only including points spaced by 2000 generations.
The results in the panel show convergence to a (global or local) maximum. Figs.   \ref{fig:gene44}(b)-(d) present the resulting optimal solution. One sees that this solution alternates between 
the same two sets of parameters as its counterpart with $\epsilon=0$. However, it does so at a finite rate that is not constrained by the lattice of time points used. The solution completes 
four full cycles in the externally mandated period $\tau$. The longer period of the cycle result in a larger variation of the probability $P_a (t)$, 
as can be seen by comparing Fig. \ref{fig:gene0}(d) with Fig. \ref{fig:gene44}(d).

A simple, but approximate, analytical description of the way that the parameter $\epsilon$ affects the period of the optimal solution can be made based on the properties of the cycles studied 
in Sec. \ref{sec:cycle}.
For small $\epsilon$ it is reasonable to expect that the optimal solution is composed of $n$ cycles with period $\tau/n$, so in each period $E_a$ and $E_b$ switch twice between
$0$ and $E_{max}$ (at different times). The barriers behave similarly. The accumulated output is therefore given by Eq. (\ref{eq:phin}). The penalty due to changes in energy is then
$- 4 n \epsilon E_{max}$. One then needs to consider the cost
\begin{equation}
C \left( \epsilon, \tau\right) = \phi_n (\tau) -  4 n \epsilon E_{max}.
\label{eq:ancost}
\end{equation}

The number of actual cycles that is expected, $\tilde{n}$, is the one that maximizes Eq. (\ref{eq:ancost}) for given $\epsilon$ and $\tau$.
It can be found by comparing the value of the cost for $n=1,2,3,\cdots$, or be approximated by finding the (non-integer)  $n^*$
that satisfies $\left. \frac{\partial C}{\partial n} \right|_{n=n^*} =0$. One then expects the genetic algorithm to converge to a solution with this number of periods. This is indeed what we observed for the cases we tested. For instance, for $\epsilon=0.044$ the cost is maximized for $n^*\simeq \tilde{n}=4$, and the results depicted in Fig. \ref{fig:gene44} indeed show a solution with
four complete cycles. Table \ref{tab:ncomp} compares the expected and observed number of cycles for several values of $\epsilon$. Excellent agreement is found between the analytical prediction and the results of the genetic algorithm in all cases.
\begin{table}

\begin{ruledtabular}
\begin{tabular}{llll}
$\epsilon $ & $n^*$ & $\tilde{n}$ & $n_{obs}$ \\
$1.21 \times 10^{-6}$ & $64.070$ & $64$ & $64$ \\
$9.70 \times 10^{-6}$ & $31.997$ & $32$ & $32$ \\
$3.39 \times 10^{-5}$ & $21.069$ & $21$ & $21$ \\
$ 7.70 \times 10^{-5}$ & $16.008$ & $16$ & $16$ \\
$6.00 \times 10^{-4}$ & $8.011$ & $8$ & $8$ \\
$1.38 \times 10^{-3}$ & $6.019$ & $6$ & $6$ \\
$4.40 \times 10^{-3}$ & $3.997$ & $4$ & $4$ \\
$9.55 \times 10^{-3}$ & $2.994$ & $3$ & $3$ \\
$2.51 \times 10^{-2}$ & $2.004$ & $2$ & $2$ \\
\end{tabular}
\end{ruledtabular}
\caption{\label{tab:ncomp} Comparison of the predicted and observed number of cycles completed in $t=\tau$ as a function of $\epsilon$. The number of cycles is controlled by a cost of changing energies, parametrised by $\epsilon$. $\tilde{n}$ is the number of compete cycles that maximizes the cost (\ref{eq:ancost}). $n^*$ is found by treating this number is being continuous rather than integer. Finally, $n_{obs}$ is the number of cycles found in the solutions identified bty the genetic algorithm.  }
\end{table}

We note that occasionally the algorithm may converge to a solution with either one additional or one missing cycle. Such solutions can serve as local maxima that trap the genetic algorithm.
While these solutions have a lower cost then the one with the correct number of cycles, the genetic algorithm sometimes finds it hard to escape from their neighborhood. Nevertheless, comparison of several independent runs of the algorithm is quite likely to find the correct maximum.

\begin{figure}[htb]
\includegraphics[scale=0.5]{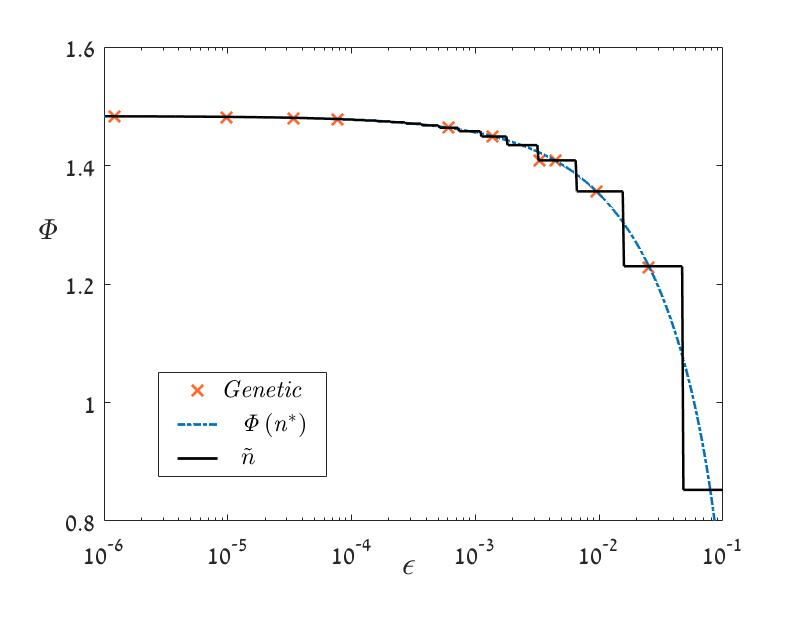}
\caption{Comparison of the output of the optimal solutions found by the genetic algorithm with the corresponding analytical estimates.}
\label{fig:comparison}
\end{figure}

Fig. \ref{fig:comparison} shows a comparison of the expected and observed output as a function of the parameter $\epsilon$.
The curves correspond to the the analytically calculated output. The blue dashed curve is obtained by substituting $n^*$ in Eq. (\ref{eq:phin}). The solid black curve is obtained by substituting the integer number of cycle that maximizes (\ref{eq:ancost}), $\tilde{n}$, in Eq.  (\ref{eq:phin}). Finally, the red crosses denotes the output of the solution obtained by the genetic algorithm.

The results depicted in Tab.  \ref{tab:ncomp} and Fig. \ref{fig:comparison} demonstrate that the analytical estimate described above matches the results of the genetic algorithm. Both the results depicted in Figs. \ref{fig:gene44} and \ref{fig:comparison} show that the optimal solutions of the model are indeed built out of the locally optimal cycles studied in Sec. \ref{sec:cycle}. As $\epsilon$ decreases
so does the period of the actual cycles that appear in the optimal solution. This reinforces the expectation that in the limit $\epsilon \rightarrow 0$ the optimal solution
will exhibit singular behavior with an infinite rate of switching between the two sets of site energies and transition barriers. After all, this singular optimal solution matches the trends seen in both the numerical and analytical approaches used here.

\section{Discussion}
\label{sec:disc}

We have studied the driving cycles that maximize the power output of a two-site stochastic pump. A combination of an analytical approach, and a numerical optimization based on a 
genetic algorithm, suggests that the optimal driving protocol switches between two system configurations in which the site energies and barriers are either maximal or minimal. Surprisingly,
the optimal driving protocol turns out to be singular, switching between the two configurations at an infinite rate. In this optimal solution, the probabilities to reside in the two sites
stay arbitrarily close to $\frac{1}{2}$. Yet, in each infinitesimally short time segment an infinitesimal amount of probability if pushed from one site to the other. Over finite times these add up
to create a finite output.

This optimal solution can be regularized by introducing an additional cost to changes in site energies. We find that this results in optimal cycles with finite periods.  Importantly, both the numerical and analytical approaches predict the same cycles. This agreement strengthens the conclusion regarding the singular
behavior of the optimal solutions when the cost assigned to energy changes approaches zero.

\begin{figure}[htb]
\includegraphics[scale=0.5]{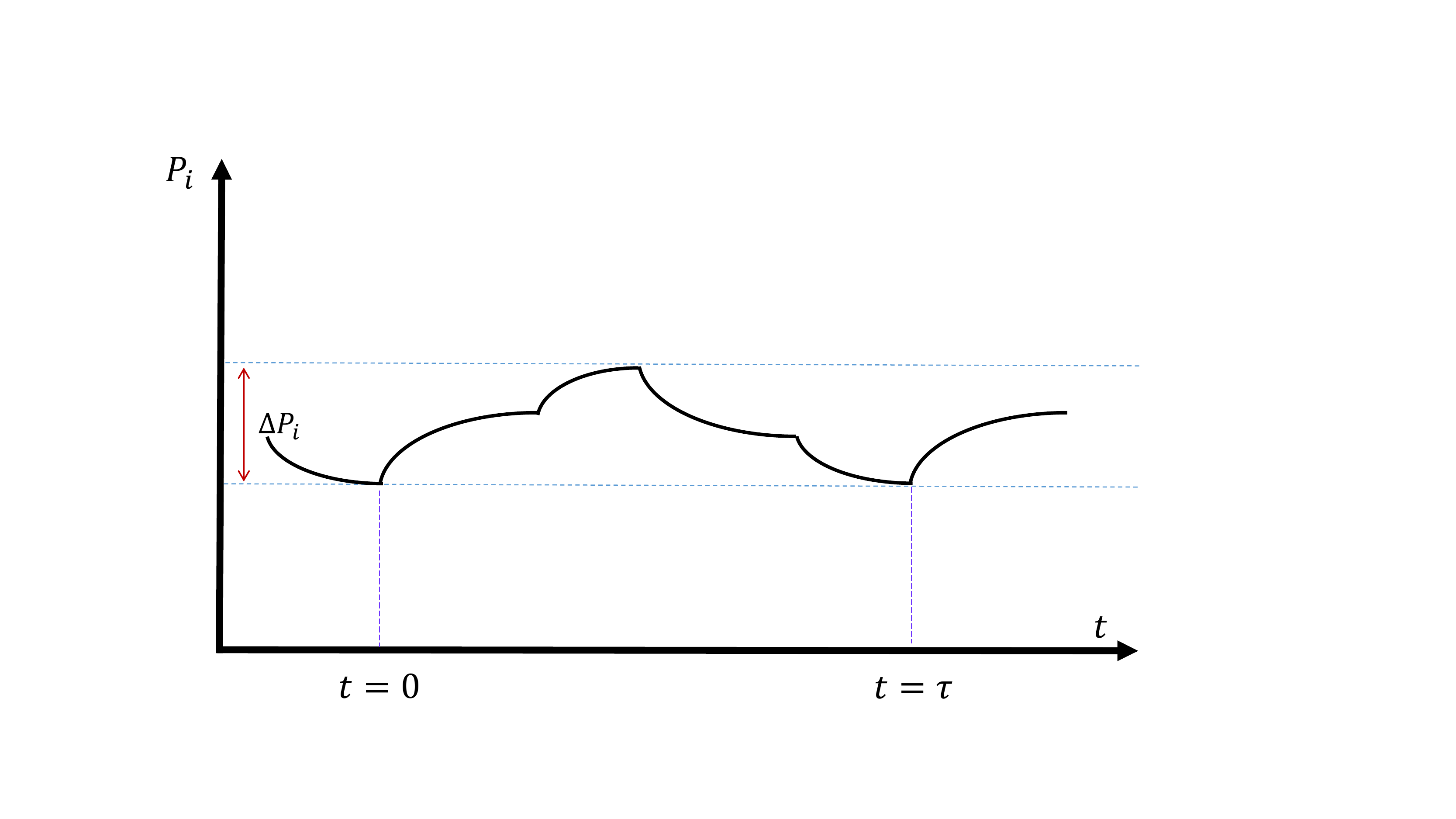}
\caption{Heuristic depiction of a cycle in a more general stochastic pump. The time dependence of the probability to be in site $i$ is shown. The cycle is composed of four different segments. In each segment the rates are assumed to be constant, and the dynamics exhibit a typical exponential relaxation towards a limiting state.}
\label{fig:hcycle}
\end{figure}

One may ask if we should have anticipated such singular optimal solutions, and how typical are they. In hindsight, it seems that such solutions should be quite typical
in overdamped systems, as long as the period is not fixed externally. Consider a similar model with several sites. For time independent rates the system follows an evolution
that qualitatively exhibits a relaxation towards a steady-state. This relaxation is often described as a sum of several exponentially decaying terms. A cycle is then typically composed of a finite chain
of such decays. Note that we assumed that any candidate for optimal solution is of the bang-bang type, composed of a series of segments in which the rates are fixed.
The probability to be in one of the sites will heuristically look like the curve depicted in Fig. \ref{fig:hcycle}.
One can expect the contribution to the output, which is generated by fluxes entering and leaving the site, to be roughly proportional to the overall change of probability in the site, $\Delta P_i$. The exact details depend on the precise configuration of allowed transitions, resisting forces, the site energies, and the barriers. By the nature of the exponential relaxation, the magnitude of the probability variation $\Delta P_i$  depends nonlinearly 
on the time spent in each segment of the cycle. Crucially, doubling the time will not result in a doubling of $\Delta P_i$.
For instance, in the two-site model studied here $\Delta P_i \propto \tanh \left( \frac{K_1+K_2}{4} \tau\right)$, see Eq. (\ref{eq:phisimple}).
But $\tanh \left( \frac{K_1+K_2}{4} \tau\right) \le \frac{K_1+K_2}{4} \tau$, meaning that the rate of accumulating output is maximized in the limit $\tau \rightarrow 0$, where the inequality is saturated. This qualitative argument suggests that such behavior 
should be quite typical for systems whose dynamics decay exponentially towards a steady-state, although one can not exclude some ways of tuning systems to achieve faster than expected relaxation \cite{Gal2020}.
All this means that singular optimal solutions with infinite rate of switching should be quite typical if the problem is set up so that infinitesimal cycle periods are allowed and make physical sense.

Finally, it is worth commenting on the possible implications of the results to the design and operation of artificial molecular machines. Indeed, an artificial molecular motor whose structure is similar to the one studied here has been realized \cite{Wilson2016}. Furthermore,
models of stochastic pumps have been
used to theoretically describe some experimental realizations of molecular machines \cite{Astumian2011}. Such models are based on an assumption of separation of time scales. Specifically, one assumes that the external
changes used to drive the molecular machines are faster compared to the time it takes the molecule to change its (mechanical) configuration. Furthermore, it is assumed that the molecule reaches 
some local equilibrium on a short time scale,  allowing one to describe its conformation changes as a memory-less (Markov) process.  Under such assumptions, the driving 
cycles used in several experiments are similar to the cycles studied in Sec. \ref{sec:cycle}. They are built of relatively fast external manipulations and time segments in which the system is 
left to evolve under constant conditions. One may wonder if such driving protocols are optimal. Should one attempt to develop more sophisticated, explicitly time-dependent 
driving mechanisms?

The results shown above suggest that such piece-wise driving protocols can be locally optimal, and that there is no reason to expect any gain from developing a driving protocol with more gradual time variation of external parameters.
The singular driving protocol that maximizes the power output breaks all the assumptions about separation of time scales described above.
However, we point out that it is not realistic to expect to be able to switch between configurations with an infinite rate. Moreover, most of the gain in power output 
is obtained when the switching is comparable to the relaxation rate. For instance, using $\tau=\frac{1}{2\left( K_1+K_2\right)}$ in (\ref{eq:phisimple}) gives about $77\%$
of the maximal output (\ref{eq:maxpower}). This means that most of the output can be achieved under conditions where  models of the type studied here are valid. 
The results presented above then suggest driving artificial molecular machines by making sudden switches between configurations at a rate that is several times faster than the typical relaxation rate of the system. 

The example studied here was of a particularly simple two-site system. It will be interesting to see how the results generalize to other, less restricted, systems. Is there a qualitative change if one breaks the symmetry between the two-sites, for instance by allowing different maximal energies? What is the optimal cycle in a system of $n$ sites linked by a cycle of transitions?
How does the optimal driving protocol look if there are two coupled cycles? Is it better to mainly drive transitions in the shorter cycle? And what happens if the maximal barrier is finite, and therefore one can not fully isolate one cycle? Such questions are left for future work.

\begin{acknowledgments}
We are grateful for support from the ISRAEL SCIENCE FOUNDATION (grant No. 1929/21). We also wish to thank Christopher Jarzynski, Sebastian Deffner, P. S. Krishnaprasad, and Christiane Koch for illuminating discussions during different stages of the research.
\end{acknowledgments}

\begin{appendix}

\section{Spectral decomposition of the rate matrix}
\label{sec:decomp}

The cycle studied in Sec. \ref{sec:cycle} was characterized by rate matrices that were piecewise constant in time. In the first half of the cycle we had
\begin{equation}
\mathbf{R}_1 = \left( \begin{matrix} -K_1 & K_2 \\ K_1 & - K_2 \end{matrix} \right).
\end{equation}
In the following we find the spectral decomposition of $\mathbf{R}_1 $, and use it to construct the propagator for the part of the cycle where $\mathbf{R}_1 $ is the rate matrix.

We start by finding the eigenvalues of $\mathbf{R}_1$.  The dynamics generated by the master equation conserve probability. As result one of the eigenvalues of $\mathbf{R}_1$ must be $\lambda_1=0$.
The other eigenvalue can be obtained from the trace of $\mathbf{R}_1 $, namely $\lambda_2=-\left( K_1+K_2\right)$.

The eigenvectors associated with the eigenvalue $\lambda_1=0$ have a simple physical interpretation. The right eigenvector is the equilibrium probability density
\begin{equation}
v_1 = \frac{1}{K_1+K_2} \left( \begin{matrix} K_2 \\ K_1\end{matrix}\right).
\end{equation}
The left eigenvector is associated with the summation over all probabilities needed for normalization,
\begin{equation}
w_1 = \left( \begin{matrix} 1 \\ 1\end{matrix}\right).
\end{equation}

The two remaining eigenvectors can be found from the demand that they satisfy bi-orthogonality relations. Specifically  $w_2^T \cdot v_1 = 0$ and $w_1^T \cdot v_2 = 0$.
One finds
\begin{equation}
v_2 = \left( \begin{matrix} 1 \\ -1\end{matrix}\right),
\end{equation}
and 
\begin{equation}
w_2 = \frac{1}{K_1+K_2} \left( \begin{matrix} K_1 \\ -K_2\end{matrix}\right).
\end{equation}

Once all the eigenvectors and eigenvalues are known, the transition rate matrix $\mathbf{R}_1 $ can be rewritten as
\begin{equation}
\mathbf{R}_1 = \lambda_1 v_1 \otimes w_1 + \lambda_2 v_2 \otimes w_2.
\label{eq:spect}
\end{equation}
Here $\otimes$ is the outer product between two vectors, given by $v \otimes w = v w^T$.

The spectral decomposition (\ref{eq:spect}) is useful because it allows us to calculate the exponential of the matrix, namely
\begin{equation}
\exp \left[\mathbf{R}_1  t \right]= e^{\lambda_1 t} v_1 \otimes w_1 + e^{\lambda_2 t}v_2 \otimes w_2= \frac{1}{K_1+K_2} \left( \begin{matrix} K_2 & K_2 \\ K_1 & K_1\end{matrix}\right)-\frac{e^{- \left( K_1+K_2\right)t}}{K_1+K_2} \mathbf{R}_1.
\end{equation}
Substitution of $t=\frac{\tau}{2}$ results in Eq. (\ref{eq:U1}).

In the second half of the cycle the transition rate matrix is given by Eq. (\ref{eq:r2}). It is easy to see that one can use the results given above also for the second half of the cycle as long as one replaces $K_1$ and $K_2$

\section{Linear response correction to the periodic state}
\label{sec:dp}

In the linear response approximation the correction to the probability distribution satisfies the inhomogeneous equation
\begin{equation}
\frac{d \delta \vec{P}}{dt} = \mathbf{R}_0 (t) \delta \vec{P} (t) + \delta \mathbf{R} (t) \vec{P}.
\end{equation}
This equation has the following formal solution
\begin{equation}
\delta \vec{P}(t) = \int_{-\infty}^t d t^\prime \mathbf{U}_0 (t,t^\prime) \delta \mathbf{R} (t^\prime) \vec{P}  (t^\prime),
\label{eq:basiclinear}
\end{equation}
where the lower limit of integration has been set so that the solution is indeed periodic in time. $\mathbf{U}_0 (t,t^\prime)$ is the propagator of the unperturbed dynamics. It satisfies
\begin{equation}
\frac{d \mathbf{U}_0 (t,t^\prime)}{dt} = \mathbf{R}_0 (t) \mathbf{U}_0 (t,t^\prime),
\end{equation}
with the initial condition $\mathbf{U}_0 (t,t)= \mathbb{I}$. 

For the two-site model, the structure of the transition rate matrix means that
\begin{equation}
\delta \mathbf{R} (t) \vec{P}  (t) = h(t) \left(\begin{matrix} 1 \\-1\end{matrix} \right),
\label{eq:pertsource}
\end{equation}
where $h(t)= \left[ \left( \delta R_{ab}^{(1)}+\delta R_{ab}^{(2)}\right) P_b  - \left( \delta R_{ba}^{(1)}+\delta R_{ba}^{(2)}\right) P_a  \right]$. Crucially, $\left( \begin{matrix} 1\\ -1 \end{matrix}\right)$ is a right eigenvector of the unperturbed transition rate matrix at all times during thr cycle. The corresponding eigenvalue is $\lambda_2=-\left( K_1+K_2\right)$. As a result,
\begin{equation}
\mathbf{U}_0 (t,t^\prime) \left( \begin{matrix} 1 \\ -1 \end{matrix} \right) = x_{t-t^\prime}  \left( \begin{matrix} 1 \\ -1 \end{matrix} \right).
\label{eq:u0one}
\end{equation}
Subsitution of Eqs. (\ref{eq:pertsource}) and (\ref{eq:u0one}) into Eq. (\ref{eq:basiclinear}) gives a simple explicit expression for the linear correction to the asymptotic periodic state
\begin{equation}
\delta \vec{P} (t) = \int_{-\infty}^t d t^\prime x_{t-t^\prime} h(t^\prime)  \left( \begin{matrix} 1 \\ -1 \end{matrix} \right).
\label{eq:tempsol}
\end{equation}
We note that Eq. (\ref{eq:tempsol}) means that $\delta P_a$ is the solution of the scalar equation
\begin{equation}
\frac{d \delta P_a}{dt} = - \left( K_1+K_2\right) \delta P_a+ h(t),
\label{eq:usfulpa}
\end{equation}
which will be used in the following.

The periodicity of $\delta \vec{P} (t) $ is ensured by the one of $h(t)$, and the simple exponential structure of $x_t$. This can be made evident by dividing the integral 
\begin{equation}
\delta \vec{P}  (t) =  \left( \begin{matrix} 1 \\ -1 \end{matrix} \right) \left[ \int_{t-\tau}^t dt^\prime x_{t-t\prime} h (t^\prime)  + \int_{t-2\tau}^{t-\tau} dt^\prime x_{t-t\prime} h (t^\prime)  + \int_{t-3\tau}^{t - 2 \tau} dt^\prime x_{t-t\prime} h (t^\prime)  + \cdots \right].
\end{equation}
Using changes of variables of the form $t_i=t^\prime + i \tau$ allows to bring all the integrals in the series back to the same domain of integration. A straightforward calculation results in
\begin{equation}
\delta \vec{P} (t) =  \left( \begin{matrix} 1 \\ -1 \end{matrix} \right) \int_{t-\tau}^t d t^\prime x_{t-t^\prime} h (t^\prime) \left[ 1+x^2+x^4 + \cdots \right] = \frac{1}{1-x^2}  \int_{t-\tau}^t d t^\prime x_{t-t^\prime} h (t^\prime)  \left( \begin{matrix} 1 \\ -1 \end{matrix} \right).
\label{eq:linearprob}
\end{equation}
This explicitly periodic expression for the correction to the periodic state, $\delta \vec{P} (t)$, is helpful in calculating the linear response correction to the power output.

Finally, some of the calculations done in the main text also require an expression for the difference $\delta {P}_a \left( \frac{\tau}{2}\right)-\delta {P}_a  (0) $.  Eq. (\ref{eq:linearprob}) allows us to write
\begin{equation}
\delta {P}_a \left( \frac{\tau}{2}\right)-\delta {P}_a  (0) = \frac{1}{1-x^2} \left[ \int_{-\frac{\tau}{2}}^{\frac{\tau}{2}} dt^\prime x_{\frac{\tau}{2}-t^\prime} h(t^\prime) - \int_0^\tau d t^\prime x_{\tau-t^\prime} h(t^\prime) \right],
\end{equation}
where in the last term we used $\delta {P}_a  (0)=\delta {P}_a  (\tau)$.
With the help of the periodicity of $h(t)$ this expression can be rewritten in terms of integrals that are limited to the range $0\le t^\prime \le \tau $. We find
\begin{equation}
\delta {P}_a  \left( \frac{\tau}{2}\right)-\delta {P}_a  (0) = \frac{1}{1+x} \left[ \int_0^{\frac{\tau}{2}} d t^\prime x_{\frac{\tau}{2}-t^\prime} h(t^\prime) - \int_{\frac{\tau}{2}}^{\tau} d t^\prime x_{\tau-t^\prime} h(t^\prime)\right].
\label{eq:dpdiff}
\end{equation}

\section{Derivation of Eq. (\ref{eq:outputstep})}
\label{sec:derivationstep}

To derive Eq. (\ref{eq:outputstep}), we note that the simple structure of the system allows us to obtain several algebraic relations between
the integrated probability fluxes
\begin{equation}
\phi_{\alpha \beta}^{(j)} \equiv \int_t^{t + \Delta t} d t^\prime R_{\alpha \beta}^{(j)} P_\beta (t^\prime).
\end{equation}
For time independent rates, there are enough such relations to solve for the fluxes and use them to calculate the output that has accumulated between $t$ and $t+\Delta t$.
Indeed, from the definition of the output it is clear that it can be recast as
\begin{equation}
\delta \Phi = \Phi (t+\Delta t)-\Phi(t)=f \left[ \phi_{ba}^{(1)}-\phi_{ab}^{(1)} -\phi_{ba}^{(2)}+\phi_{ab}^{(2)}\right].
\label{eq:dphi}
\end{equation}

It is straightforward to relate the change in probabilities to a linear combination of the fluxes
\begin{equation}
P_a(t+\Delta t) -P_a (t) = P_b (t) - P_b (t+\Delta t) = \phi_{ba}^{(1)}+\phi_{ba}^{(2)} -\phi_{ab}^{(1)} - \phi_{ab}^{(2)}.
\label{eq:fluxprob}
\end{equation}
For time independent transition rates one also finds
\begin{equation}
 \frac{\phi_{ba}^{(1)}}{\phi_{ba}^{(2)}} = \frac{R_{ba}^{(1)}}{R_{ba}^{(2)}},
\end{equation}
and
\begin{equation}
 \frac{\phi_{ab}^{(1)}}{\phi_{ab}^{(2)}} = \frac{R_{ab}^{(1)}}{R_{ab}^{(2)}}.
\end{equation}
Finally, one can use conservation of probability $P_b (t)+P_a(t)=1$ to write e.g.
\begin{equation}
\frac{\phi_{ab}^{(1)}}{R_{ab}^{(1)}} +\frac{\phi_{ba}^{(2)}}{R_{ba}^{(2)}} = \Delta t.
\label{eq:conserve}
\end{equation}

Equations (\ref{eq:fluxprob})-(\ref{eq:conserve}) can be solved to give
\begin{eqnarray}
\phi_{ab}^{(1)} & = & - \frac{1}{\left| \lambda_2 \right|} R_{ab}^{(1)} \Delta P_b + \frac{1}{\left| \lambda_2 \right|} \left( R_{ba}^{(1)} +R_{ba}^{(2)}\right) R_{ab}^{(1)} \Delta t, \nonumber \\  
\phi_{ba}^{(1)} & = & + \frac{1}{\left| \lambda_2 \right|} R_{ba}^{(1)} \Delta P_b + \frac{1}{\left| \lambda_2 \right|} \left( R_{ab}^{(1)} +R_{ab}^{(2)}\right) R_{ba}^{(1)} \Delta t, \nonumber \\
\phi_{ab}^{(2)} & = & - \frac{1}{\left| \lambda_2 \right|} R_{ab}^{(2)} \Delta P_b + \frac{1}{\left| \lambda_2 \right|} \left( R_{ba}^{(1)} +R_{ba}^{(2)}\right) R_{ab}^{(2)} \Delta t, \nonumber \\  
\phi_{ba}^{(2)} & = & + \frac{1}{\left| \lambda_2 \right|} R_{ba}^{(2)} \Delta P_b + \frac{1}{\left| \lambda_2 \right|} \left( R_{ab}^{(1)} +R_{ab}^{(2)}\right) R_{ba}^{(2)} \Delta t.
\label{eq:phis}
\end{eqnarray}
As before, the eigenvalue associated with the relaxation rate is $\lambda_2 = - R_{ab}^{(1)} -R_{ab}^{(2)} - R_{ba}^{(1)} -R_{ba}^{(2)}$. Substitution of Eqs. (\ref{eq:phis}) in 
(\ref{eq:dphi}) leads to Eq. (\ref{eq:outputstep}).
\end{appendix}

\end{document}